\documentclass[10pt,twocolumn]{aastex63}

\def\chandra{{\itshape Chandra\/}}
\def\xmm{{\itshape XMM-Newton\/}}
\def\nustar{{\itshape NuSTAR\/}}

\def\xray{\hbox{X-ray}}

\def\ltsima{$\; \buildrel < \over \sim \;$}
\def\simlt{\lower.5ex\hbox{\ltsima}}
\def\gtsima{$\; \buildrel > \over \sim \;$}
\def\simgt{\lower.5ex\hbox{\gtsima}}

\usepackage{apjfonts}
\usepackage{xcolor}
\usepackage{mwe}
\usepackage{lipsum}
\usepackage[shortcuts]{extdash}

\begin{document}

\title{The Stellar Age Dependence of X-ray Emission from Normal Star-Forming Galaxies in the GOODS Fields}

\correspondingauthor{Woodrow Gilbertson}
\email{wagilber@uark.edu}

\author[0000-0002-4049-5314]{Woodrow Gilbertson}
\affiliation{Department of Physics, University of Arkansas, 226 Physics Building, 825 West Dickson Street, Fayetteville, AR 72701, USA}

\author[0000-0003-2192-3296]{Bret D. Lehmer}
\affiliation{Department of Physics, University of Arkansas, 226 Physics Building, 825 West Dickson Street, Fayetteville, AR 72701, USA}
\affiliation{Arkansas Center for Space and Planetary Sciences, University of Arkansas, 332 N. Arkansas Avenue, Fayetteville, AR 72701, USA}

\author[0000-0001-5035-4016]{Keith Doore}
\affiliation{Department of Physics, University of Arkansas, 226 Physics Building, 825 West Dickson Street, Fayetteville, AR 72701, USA}
\affiliation{Arkansas Center for Space and Planetary Sciences, University of Arkansas, 332 N. Arkansas Avenue, Fayetteville, AR 72701, USA}

\author[0000-0002-2987-1796]{Rafael T. Eufrasio}
\affiliation{Department of Physics, University of Arkansas, 226 Physics Building, 825 West Dickson Street, Fayetteville, AR 72701, USA}

\author[0000-0001-8525-4920]{Antara Basu-Zych}
\affiliation{Department of Physics, University of Maryland Baltimore County, Baltimore, MD 21250, USA}
\affiliation{NASA Goddard Space Flight Center, Laboratory for X-ray Astrophysics, Greenbelt, MD 20771, USA}
\affiliation{Center for Research and Exploration in Space Science and Technology, NASA/GSFC, Greenbelt, MD 20771}

\author[0000-0002-0167-2453]{William N. Brandt}
\affiliation{Department of Astronomy and Astrophysics, 525 Davey Lab, The Pennsylvania State University, University Park, PA 16802, USA}
\affiliation{Institute for Gravitation and the Cosmos, The Pennsylvania State University, University Park, PA 16802, USA}
\affiliation{Department of Physics, 104 Davey Laboratory, The Pennsylvania State University, University Park, PA 16802, USA}

\author[0000-0003-1474-1523]{Tassos Fragos}
\affiliation{Departement d’Astronomie, Universite de Geneve, Chemin Pegasi 51, CH-1290 Versoix, Switzerland}

\author[0000-0002-9202-8689]{Kristen Garofali}
\affiliation{Department of Physics, University of Arkansas, 226 Physics Building, 825 West Dickson Street, Fayetteville, AR 72701, USA}
\affiliation{NASA Goddard Space Flight Center, Laboratory for X-ray Astrophysics, Greenbelt, MD 20771, USA}

\author[0000-0003-3684-964X]{Konstantinos Kovlakas}
\affiliation{Departement d’Astronomie, Universite de Geneve, Chemin Pegasi 51, CH-1290 Versoix, Switzerland}

\author[0000-0002-9036-0063]{Bin Luo}
\affiliation{School of Astronomy and Space Science, Nanjing University, Nanjing, Jiangsu 210093, China}
\affiliation{Key Laboratory of Modern Astronomy and Astrophysics (Nanjing University), Ministry of Education, Nanjing, Jiangsu 210093, China}

\author[0000-0003-3096-9966]{Paolo Tozzi}
\affiliation{INAF, Osservatorio Astrofisico di Firenze, Largo Enrico Fermi 5, I-50125, Firenze, Italy}

\author[0000-0003-0680-9305]{Fabio Vito}
\affiliation{Scuola Normale Superiore, Piazza dei Cavalieri 7, 56126 Pisa, Italy}

\author[0000-0002-7502-0597]{Benjamin F. Williams}
\affiliation{Department of Astronomy, Box 351580, University of Washington, Seattle, WA 98195, USA}

\author[0000-0002-1935-8104]{Yongquan Xue}
\affiliation{CAS Key Laboratory for Research in Galaxies and Cosmology, Department of Astronomy, University of Science and Technology of China, Hefei 230026, China}
\affiliation{School of Astronomy and Space Sciences, University of Science and Technology of China, Hefei 230026, China}

\begin{abstract}

The \textit{Chandra} Deep Field-South and North surveys (CDFs) provide unique windows into the cosmic history of X-ray emission from normal (non-active) galaxies. Scaling relations of normal galaxy X-ray luminosity ($L_{\rm{X}}$) with star formation rate (SFR) and stellar mass ($M_\star$) have been used to show that the formation rates of low-mass and high-mass X-ray binaries (LMXBs and HMXBs, respectively) evolve with redshift across $z \approx$~0\==2 following $L_{\rm{HMXB}} / \rm{SFR} \propto (1 + \emph{z})$ and $L_{\rm{LMXB}} / M_\star \propto (1 + z)^{2-3}$. However, these measurements alone do not directly reveal the physical mechanisms behind the redshift evolution of X-ray binaries (XRBs). We derive star-formation histories for a sample of 344 normal galaxies in the CDFs, using spectral energy distribution (SED) fitting of FUV-to-FIR photometric data, and construct a self-consistent, age-dependent model of the X-ray emission from the galaxies. Our model quantifies how \xray\ emission from hot gas and XRB populations vary as functions of host stellar-population age. We find that (1) the ratio $L_{\rm{X}}/M_\star$ declines by a factor of $\sim$1000 from 0\==10 Gyr and (2) the X-ray SED becomes harder with increasing age, consistent with a scenario in which the hot gas contribution to the X-ray SED declines quickly for ages above 10~Myr. When dividing our sample into subsets based on metallicity, we find some indication that $L_{\rm{X}}/M_\star$ is elevated for low-metallicity galaxies, consistent with recent studies of X-ray scaling relations. However, additional statistical constraints are required to quantify both the age and metallicity dependence of X-ray emission from star-forming galaxies.

\end{abstract}


\section{Introduction} \label{sec:intro}

With the advent of deep extragalactic \xray\ surveys with \chandra, such as the \chandra\ Deep Fields (CDFs; \citealp{xue,luo,Xue_2017}), it has become possible to perform statistically meaningful studies of the \xray\ emission from cosmologically distant ($z \approx$~0.1\==3) normal (non-active) galaxies. Detailed studies of normal galaxies in the local universe have shown that hot gas and \xray\ binary populations (XRBs) are responsible for the majority of the $\approx$0.3\==1~keV and $\approx$1\==30~keV emission, respectively (see e.g., \citealp{mineo_12a, mineo_12b,lehmer_15,garofali_20}). Due to the redshifting of rest-frame emission, combined with the \chandra\ response peak at $\approx$1~keV, studies of cosmologically distant galaxies have provided important empirical insight into the evolution of XRB populations with cosmic time (see, e.g., \citealp{Norman_2004,Brandt_2005,nandra,lehmer_05,lehmer_07,lehmer_08,lehmer_16,ptak,Basu_Zych_2013a,mineo_14,aird_16,fornasini_19,fornasini_20}). \\ \indent 
It has been known for some time that the collective \xray\ emission from populations of relatively young high-mass XRBs (HMXBs) and old low-mass XRBs (LMXBs) each scale with star-formation rate (SFR) and stellar mass ($M_\star$) respectively; hereafter, the $L_{\rm X}$(HMXB)/SFR and $L_{\rm X}$(LMXB)/$M_\star$ scaling relations (see, e.g., \citealp{grimm,ranalli_03,gilfanov_04,fabbiano_06,lehmer_10,mineo_12b,zhang,fabbiano_19,lehmer_19,lehmer_21}). Using the {\ttfamily Millenium~II} cosmological simulation with a prescription for baryon evolution, along with the {\ttfamily Startrack} binary population synthesis code, \cite{fragos_13a} identified a set of population synthesis models consistent with the local XRB scaling relations (\citealp{Boylan_Kolchin_2009,Guo_2011,Belczynski_2008}). In the process, these models predicted that the \xray\ scaling relations should increase substantially with redshift as a result of declining chemical abundances (metallicities) and stellar ages; two factors that are expected to have a significant impact on XRB population emission. Stellar age, the main focus of this paper, is expected to play a role as the average masses of donor stars in LMXBs decreases with age, leading to decreased \xray\ luminosity (e.g. \citealp{fragos_13a,fragos_13b,lehmer_14}). \\ \indent
\cite{lehmer_16} performed stacking analyses of galaxy populations in the CDFs and measured scaling relations over $z \approx$~0\==2, which could then be used to empirically test the \citet{fragos_13a} predictions. They found that the XRB scaling relations indeed appear to undergo substantial evolution, with best-fit relations $L_{\rm X}$(HMXB)/SFR~$\propto (1+z)$ and $L_{\rm X}$(LMXB)/$M_\star \propto (1+z)^{2-3}$ over this redshift range. Similar results were obtained by \cite{aird_17} based on the CDFs and \chandra\ COSMOS surveys. The empirical constraints on the redshift evolution of the scaling relations were consistent with some of the \cite{fragos_13a} models, indirectly supporting the conclusions that metallicity and stellar age are major factors impacting XRB scaling relations and further constraining some of the model parameters in population synthesis studies. \\ \indent
More recently, attention has focused on more direct empirical tests of how physical properties, like metallicity and star-formation history, impact the formation frequency of XRBs. For relatively nearby ($D\simlt 50$~Mpc) actively star-forming galaxies that have high specific-SFR (sSFR, $\equiv$SFR/$M_\star$), investigations have shown that the number of bright HMXBs (e.g., ultraluminous \xray\ sources [ULXs]) and the HMXB power output per unit SFR ($N_{\rm ULX}$/SFR and $L_{\rm X}$(HMXB)/SFR, respectively) appear to increase at low metallicity (see, e.g., \citealp{mapelli,Basu_Zych_2013b,basu_16,prestwich,brorby_14,brorby_16,douna,kovlakas_20,lehmer_21,Saxena_2021}), consistent with population synthesis predictions (e.g., \citealp{linden_10,fragos_13a,fragos_13b,wiktorowicz_17,wiktorowicz_19}). For a subset of actively star-forming galaxies at $z \approx$~0.1\==2.6 in the CDFs and COSMOS fields that have reliable metallicity estimates from spectroscopy, \cite{fornasini_19,fornasini_20} showed that the $L_{\rm X}$(HMXB)/SFR versus metallicity relation for distant galaxies is consistent with that of local galaxies. These works provide strong evidence that the observed redshift evolution of $L_{\rm X}$(HMXB)/SFR is driven by metallicity evolution. The suggested mechanism for this evolution is that lower-metallicity stars have weaker stellar winds, resulting in lower mass-loss over their lifetimes. This leads to more massive compact objects, which in turn leads to a higher quantity of XRBs, as well as more luminous HMXBs per unit SFR.\\ \indent
Empirically constraining the age evolution of XRB populations over $\approx$10~Myr to $\approx$10~Gyr timescales has been relatively challenging. Targeted observations of early-type galaxies with stellar-population ages that span $\approx$3\==12~Gyr (e.g., \citealp{kim_10,lehmer_14}) have provided tantalizing evidence for variations in the $L_{\rm X}$(LMXB)/$M_\star$ scaling relation, in line with those predicted by population synthesis models (however, see \citealp{lehmer_20}). Also, recent investigations of XRB formation rates within subgalactic regions of nearby galaxies (e.g., the Magellanic Clouds, M33, M51, NGC~3310, and NGC~2276) have revealed preferred timescales for XRB formation consistent with basic theoretical expectations: HMXBs forming <100 Myr after an initial star-formation event and rapidly declining after that, with LMXBs maintaining a much flatter power distribution out to Gyr timescales before breaking and declining (e.g., \citealp{antoniou_16,lehmer_17,garofali_18,anastasopoulou_18,antoniou_19,lazzarini_21}). Nonetheless, a detailed framework for how XRB populations evolve with age has yet to be rigorously tested. \\ \indent
In this paper, we make use of the plethora of UV\==to\==far-IR data in the Great Observatories Origins Deep Survey (GOODS; \citealt{giavalisco_04}), in concert with the CDF \xray\ data, to sensitively determine the star-formation histories (SFHs) and their correlation to the X-ray emissions in normal galaxies. We build upon the statistical techniques developed by \cite{lehmer_17}, and use detailed information about galaxy SFHs, along with \chandra\ constraints on galaxy \xray\ emission, to construct a self-consistent model for hot gas and XRB emission per unit mass as a function of age. \\ \indent
In Section \ref{sec:sample_select}, we outline the galaxy sample selection process, and our efforts to eliminate possible active galactic nuclei (AGN). In Section \ref{sec:derivation}, we discuss the process of fitting the UV\==to\==IR spectral energy distributions of the galaxies in our sample to estimate SFHs. We present physical and \xray\ properties of the sample such as stellar mass, star-formation history, redshift, metallicity, and \xray\ counts. In Section \ref{sec:tech}, we introduce our models to directly constrain the age-dependence of $L_{\rm X}/M_\star$ and the \xray\ spectral shape. The statistical methods used are discussed and the results from the parameter fitting are presented. In Section \ref{sec:discussion}, we interpret and parameterize our results, test them for biases, and assess whether additional physical properties influence \xray\ emission from the galaxies. \\ \indent
Throughout this paper, we make use of the main point-source catalogs and data products for the $\approx$2~Ms CDF-N and $\approx$7~Ms CDF-S as outlined in \cite{xue} and \cite{luo}, respectively (see \citealp{Brandt_2015} for a review on CDF surveys). The Galactic column densities are $1.6 \times 10^{20}$~cm$^{-2}$ and $8.8 \times 10^{19}$~cm$^{-2}$ for the CDF-N and CDF-S, respectively \citep{stark}. All of the \hbox{X-ray} fluxes and luminosities quoted throughout this paper have been corrected for Galactic absorption. Estimates of $M_\star$, SFR, and SFH presented throughout this paper were derived assuming a \cite{kroupa_01} initial mass function (IMF); when making comparisons with other studies, we have adjusted all values to correspond to our adopted IMF. Cosmological values of $H_0$ = 70~\hbox{km s$^{-1}$ Mpc$^{-1}$}, $\Omega_{\rm M}$ = 0.3, and $\Omega_{\Lambda}$ = 0.7 are adopted throughout this paper. \\
\begin{figure*}
	\centering
	\begin{minipage}{8.75cm}
	\centering
	\includegraphics[width=1\textwidth]{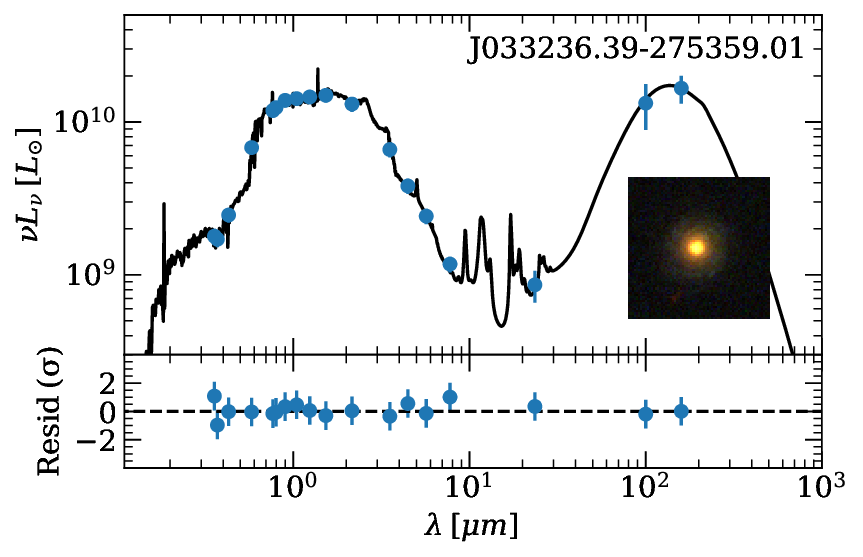}
	\end{minipage} \hfill
    \begin{minipage}{8.75cm}
    \centering
	\includegraphics[width=1\textwidth]{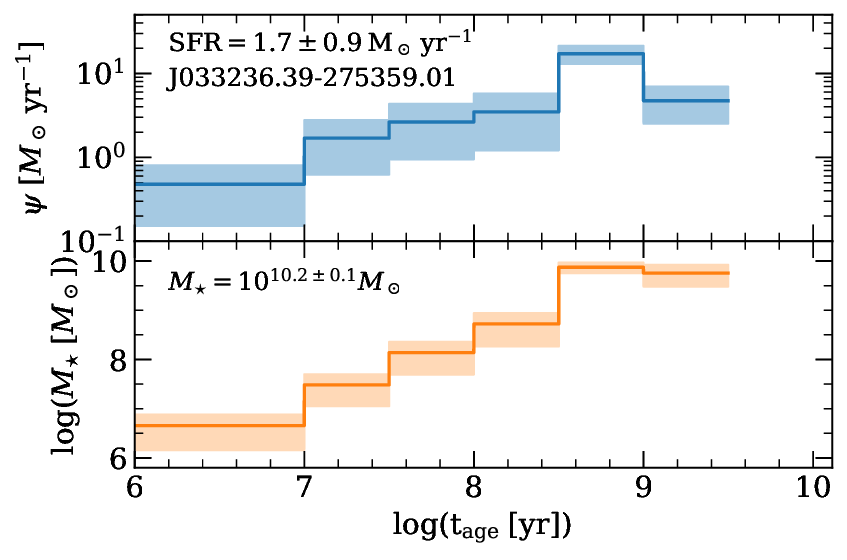}
	\end{minipage}
\caption{\textit{Left, Top}: SED fitting results for J033236.39$-$275359.01, a $z =$~0.52 galaxy in our CDF-S sample. The blue filled circles and $1\sigma$ error bars represent observational data. The black line shows the best-fit model SED from {\ttfamily Lightning} (see Section~\ref{sub:phys_deriv}). The inset three-color, 16 square arcsec optical image shows J033236.39$-$275359.01 as viewed by {\it HST} ({\it red}=F850LP, {\it green}=F606W, and {\it blue}=F435W). \textit{Left, Bottom}: Residuals to the SED fit, calculated as the data$-$model, in units of $\sigma$. The dashed black line shows zero, for reference. \textit{Right}: The resulting SFH from our SED fitting of J033236.39$-$275359.01, represented in terms of SFR as a function of time ({\it top}) and contributions to the current stellar-mass as a function of population age ({\it bottom}). The SFR listed in the upper left is taken as an average over the last 100 Myr. Shaded regions show 1$\sigma$-level uncertainties. The apparent rise in stellar mass contributions with increasing population age is driven primarily by the sizes of the age bins growing with increasing age. The 7th age bin (3.16\==10~Gyr) is fixed at zero for this galaxy, due to the age of the Universe being $\approx$8.3~Gyr at $z = 0.52$.
}
\label{fig:SED}
\end{figure*}
\section{Sample Selection} \label{sec:sample_select}
Given that the goal of this study is to examine how \hbox{X-ray} emission scales with galaxy properties, it is essential that we construct a sample with properties that are well determined. We began by constructing a sample of galaxies in the GOODS-North and South fields; hereafter, the GOODS-N and GOODS-S, respectively. The extensive multiwavelength coverage in the GOODS fields allows for high-quality constraints on galaxy SFHs through spectral energy distribution (SED) fitting from the UV-to-IR (see Section~\ref{sub:phys_deriv} for details). \\ \indent
We chose to utilize the GOODS-N catalog from \cite{2019ApJS..243...22B} and GOODS-S catalog from \cite{2013ApJS..207...24G}, which contain 35,445 and 34,930 sources, respectively, for a total of 70,375 initial objects\footnote{Retrieved from the \textit{Rainbow} database: \url{http://rainbowx.fis.ucm.es/Rainbow_navigator_public/}}. We restricted our sample to include only galaxies that had 6 or more detections at wavelengths greater than or equal to 3.6~$\mu$m as to better constrain the dust emission SED (see Section \ref{sec:derivation} for details). This restriction required detections in all \textit{Spitzer}/IRAC bands and two or more detections in any combination of the \textit{Spitzer}/MIPS 24~$\mu$m and 70~$\mu$m, \textit{Herschel}/PACS 100~$\mu$m and 160~$\mu$m, and \textit{Herschel}/SPIRE 250~$\mu$m bands, none of which suffered from photometric blending effects significant enough to differentiate them from super-deblended catalogs, such as \cite{Liu_2018}. Detections at 350~$\mu$m and 500~$\mu$m from SPIRE were ignored due to significant potential blending issues (\citealp{doore_2021}). \\ \indent
We also required that galaxies in our sample have ``good'' spectroscopic redshifts (as defined by having no warning flag in the data) which were compiled from numerous other studies and used for source identification \citep{2004AJ....127.3121W, 2004ApJS..155..271S, 2005A&A...437..883M, 2006ApJ...653.1004R, 2007A&A...465.1099R, 2008A&A...478...83V, 2008ApJ...689..687B, 2009A&A...494..443P, 2010A&A...512A..12B, 2010ApJ...719..425F, 2011AJ....141....1T, 2012MNRAS.425.2116C, 2015ApJS..218...15K}. With good spectroscopic redshifts, we can confidently calculate distance measurements as accurately as our cosmology allows, and assure that the errors calculating rest-frame quantities (e.g., SFHs and X-ray luminosities) are minimal. The above requirements yielded a sample of 674 galaxies in the GOODS-N and 430 galaxies in the GOODS-S, for a total of 1,104 galaxies. \\ \indent
We further applied AGN-related cuts to limit our sample to ``normal'' galaxies that have \xray\ emission dominated by hot gas and XRBs. To begin with, we adopted the \xray\ source classifications provided by \cite{xue} and \cite{luo} for the CDF-N and CDF-S, respectively, which make use of total $L_{\rm{X}}$, photon index ($\Gamma$), \xray-to-optical/IR/radio flux ratio ($f_{\rm X}/f_{\rm opt}$), and optical broad emission-line features to characterize sources as AGN, galaxies, and foreground stars. We also searched the \cite{Ding_2018} catalog of variability-selected AGN in the CDF-S, and found that all such AGN had already been removed by other sample selection criteria. The AGN candidate sources were located using the maximum-likelihood optical counterpart coordinates from the Cosmic Assembly Near-infrared Deep Extragalactic Legacy Survey (CANDELS; \citealp{Grogin_2011,koekemoer_2011}), which are provided in the \cite{xue} and \cite{luo} catalogs. These CANDELS coordinates were also provided with the data used for SED fitting, allowing for a straightforward matching procedure. A matching radius of 0.5 arcsec was used across all sources, as the only difference between source locations came from rounding/precision differences. Any source classified as an AGN by the respective \xray\ catalog, or found within 10 arcsec of a source flagged as an AGN, was removed from the sample. This radius was selected to robustly reduce as much as AGN contamination as possible. While these criteria optimally divided X-ray sources into appropriate classifications, there are inevitably some sources that are truly AGN that are classified as galaxies (and vice-versa) near the classification boundaries. \\ \indent
To more conservatively remove AGN from our sample, we subsequently adopted more stringent limits than those published and excluded sources with $\log(f_{\rm X}/f_{\rm F850LP}) > -3$ and a 2\==7~keV to 0.5\==2~keV count rate ratio (hereafter, band ratio) BR ~$>0.6$ (comparable to an effective power-law slope of $\Gamma_{\rm eff} \le 1.5$) as potential AGN contaminants (compared with $\log(f_{\rm X}/f_{R}) > -1$ and $\Gamma_{\rm eff} \le 1$ adopted by \citealp{xue,luo}). We expect that these stringent selection criteria will result in the removal of mainly normal galaxies (particularly those with harder \xray\ emission); however, there are known AGN with properties in this range and we therefore flag such sources as plausible AGN. These cuts specifically removed 36 galaxies in the CDF-S and 10 galaxies in the CDF-N. In Section~\ref{sub:subs}, we show that this more conservative AGN selection does not have any material impact on our results. Additionally, we checked our sample for AGN with the WISE-magnitude system optimized for completeness presented in \cite{assef_2018} and found that none of our sources were flagged as AGN using the given criteria ($W1 - W2 > 0.77$). After the above cuts were applied, we were left with 253 galaxies in the GOODS-N and 214 galaxies in the GOODS-S, for a sample size of 467 galaxies. \\ \indent
The final cut applied to the sample was to limit the off-axis angle of galaxies. The \textit{Chandra} coordinates were used, limiting our sample to galaxies within 6 arcmin of the average aim point of each field (\citealp{xue,luo}). This limit was applied to ensure that the sources were located in the most sensitive regions of the CDFs where the \textit{Chandra} point spread function was small and well behaved. This resulted in 200 galaxies in the GOODS-N and 144 galaxies in the GOODS-S, and thus, a total sample size of 344 galaxies. Throughout the remainder of this paper we will be using this sample of 344 galaxies, and will be quoting results as we fit models to this sample. \\ \indent
\begin{figure}
	\centering
	\includegraphics[width=8.75cm]{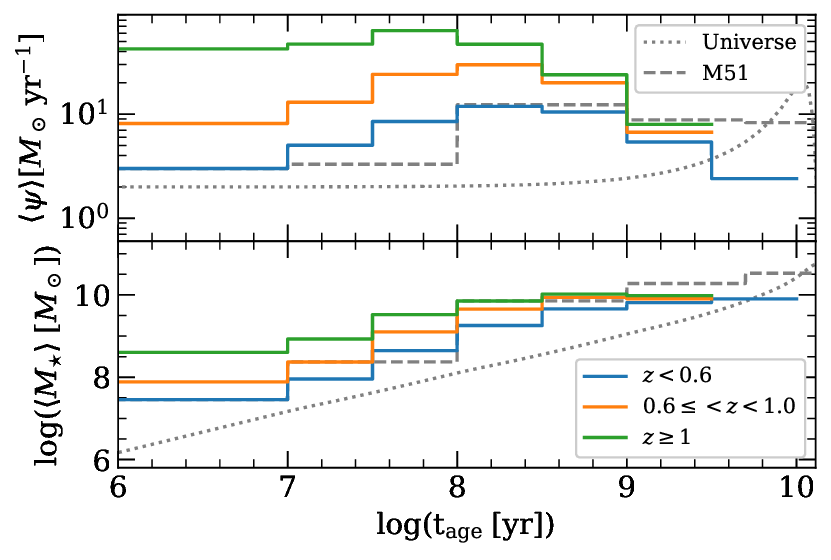}
	\caption{The mean SFHs of galaxy subsamples, expressed in terms of star-formation rates, $\psi$ ({\it top}), and age-dependent contributions to stellar masses ({\it bottom}). The SFHs were created by fitting the SEDs of all galaxies in subsamples broken up into redshift bins. The solid blue lines show 119 galaxies at $z < 0.6$. The solid orange lines show 123 galaxies in the range of $0.6 \leq z < 1.0$. These redshift bins are only displayed here and not used for analysis throughout the paper, and demonstrated to be unimportant to the model in Section \ref{sub:subs}. The solid green lines show 102 galaxies at $z \geq 1.0$. The dashed gray line shows the SFH of M51 \citep{eufrasio}, which was studied by \citet{lehmer_17} to constrain the age-evolution of XRB populations and is used as a comparison in this study. The dotted gray line shows the model for the average SFR density of the Universe (\citealp{madau_17}) with an arbitrary normalization applied such that the mass and SFR of the 1$-$3.16 Gyr population is comparable to those in this study.}
	\label{fig:SFH_all}
\end{figure}
\section{Sample Properties} \label{sec:derivation}
\subsection{Derivation of Physical Properties} \label{sub:phys_deriv}
To study how the X-ray emission from normal galaxies evolves with age, we began by estimating the SFH of each galaxy using the SED-fitting code {\ttfamily Lightning}\footnote{Version 2.0 \url{https://github.com/rafaeleufrasio/lightning}} \citep[][]{eufrasio,doore_2021}. {\ttfamily Lightning} models the overall UV-to-IR SED of a given galaxy using SED contributions from stellar populations within distinct age bins, which are evenly spaced in log space, (e.g. 0\==10 Myr, 10\==31.6 Myr, 31.6\==100 Myr, 0.1\==0.316 Gyr, 0.316\==1 Gyr, 1\==3.16 Gyr, and 3.16\==10 Gyr for this study, listed in Table \ref{table:agebins_coefs}), and includes dust attenuation and emission prescriptions that are tied together via energy balance. While not included in this study, we did test processes using 5 and 10 age-bins, and found that the results are largely insensitive to the number of age-bins. {\ttfamily Lightning} employs a Markov Chain Monte Carlo (MCMC) algorithm and provides SFH posteriors in the form of MCMC chains. \\ \indent
\begin{table}
\centering
 \begin{tabular}{ l r }
 \hline
 \hline
 Age bin range & $M_\star/M_0$ \\
 \hline
 0--10 Myr & 0.947 \\
 10--31.6 Myr & 0.832 \\
 31.6--100 Myr & 0.764 \\
 100--316 Myr & 0.700 \\
 0.316--1 Gyr & 0.636 \\
 1--3.16 Gyr & 0.555 \\
 3.16--10 Gyr & 0.484 \\
 \hline
 \hline
\end{tabular}
\caption{The age ranges for each of the seven bins used throughout this study, as well as the fraction of current stellar mass ($M_\star$) over initial stellar mass ($M_0$) used to convert from SFR to current stellar mass.}
\label{table:agebins_coefs}
\end{table}
Since our sample spans a significant range of redshifts ($z =$~0\==3.5), the inclusion of the two oldest age bins for a given galaxy depends on the age of the Universe at that galaxy's redshift. When fitting the SED of a given galaxy, we required that all age bins have upper age bounds less than the age of the Universe for inclusion in our SED modeling. For $z>0.3$ ($z>2$) galaxies, where the age of the Universe was less than 10 Gyr (3.2~Gyr), the SFR across the 3.16\==10 Gyr (1\==3.16~Gyr) age bin was set to zero. We chose to set these age bins to zero to prevent the SED model from including stars of ages older than the Universe. In practice, there will be some emission from stars in these old age bins, and our modeling procedure forces this emission to be attributed to stars in younger bins, potentially elevating the stellar mass associated with the young population.  However, our fits to the SEDs are already good in a statistical sense (via null-hypothesis probabilities) without the inclusion of these old populations, suggesting that the impact of this choice on the stellar masses of the younger populations that are modeled are minimal, and below the level of the uncertainties. \\ \indent
The SED model for each stellar-age range is based on the spectrum produced by stellar populations formed across the age range assuming a constant SFR over the bin. As such, our SFHs can be expressed as either the SFR in each age-bin, $\psi_i$, or the stellar-mass contributions from populations for the seven distinct age ranges $M_{\star,i}$. \\ \indent
To model the SEDs of each galaxy, we utilized the \citet{2000ApJ...533..682C} attenuation curve and dust\==emission models from \citet{2007ApJ...657..810D}. The \citet{2000ApJ...533..682C} attenuation curve has a single free parameter, the optical depth of the diffuse dust in the \textit{V}-band ($\tau_V^{\rm diff}$), which is used for normalization and is proportional to $A_V$ ($A_V=1.086\tau_V^{\rm diff}$). For the \citet{2007ApJ...657..810D} dust\==emission model, there are five free parameters ($\alpha$, $U_{\rm min}$, $U_{\rm max}$, $\gamma$, $q_{\rm PAH}$). The dust emission normalization parameter is uniquely determined by the attenuated UV/optical power in the energy balance assumption. Of these five dust parameters, we fix $U_{\rm max} = 3 \times 10^5$ and $\alpha = 2$, motivated by the work of \citet{2007ApJ...663..866D} who found that fits to the IR dust emission using the \citet{2007ApJ...657..810D} model were not sensitive to precise values of $U_{\rm max}$ and $\alpha$. Further, they found that the values of $U_{\rm max} = 10^6$ and $\alpha = 2$ could produce quality fits to IR SEDs. Thus, fixing these values minimizes any possible degeneracies that might arise from the insensitivity of these parameters.\footnote{We note that our fixed value of $U_{\rm max}$ is different from that found in \citet{2007ApJ...663..866D}. This is due to \texttt{Lightning} not extrapolating the publicly available $\delta$-functions of $U$ from \citet{2007ApJ...657..810D} \citep[see][for more details]{doore_2021}.} We also limit the dust models to use the ``restricted'' form ($0.7 \le U_{\rm min} \le 25$) which is recommended by \citet{2007ApJ...663..866D} when submillimeter data is unavailable. \\ \indent
Using our chosen SFH age bins, attenuation curve, and dust\==emission model; we fit the SED of each galaxy assuming flat priors on all free parameters. To test the quality of the resulting fits, we performed a $\chi^2$ goodness-of-fit test on each galaxy, which indicates acceptable fits with no excessive over- or under-fitting. Therefore, our resulting SFH estimates and corresponding uncertainties for these galaxies are statistically sound. \\ \indent
In Figure \ref{fig:SED}, we show the SED data quality and best-fit {\ttfamily Lightning} model for J033236.39$-$275359.01, an example $z = $~0.52 galaxy in the CDF-S. The observed photometry and modeled SED are shown on the left, along with the residuals. On the right, we display the resulting SFH and $1\sigma$ uncertainties in terms of $\psi_i$ and $M_{\star,i}$ in the seven age ranges. As we outline below, the $M_{\star,i}$ representation of the SFH is essential for this study, as we aim to derive how the X-ray SED per unit stellar mass evolves as a function of age. \\ \indent
Using our SFHs, we calculated the rest-frame SFR and $M_\star$ for our galaxies. For consistency with other studies, we define and calculate rest-frame $\rm{SFR}$ as the mean value of $\psi_i$ over the last 100 Myr, which is calculated for each galaxy as the age-weighted average of $\psi_i$ values from the three youngest age bins. As for $M_{\star}$, this is the total stellar mass calculated by converting the SFR of each SFH age bin to mass and summing over all age bins. The conversion factors are a factor of width of the age bin and the fraction of stars surviving from that age bin, listed in Table \ref{table:agebins_coefs}. These factors include the effects of stellar evolution. \\ \indent
In Table \ref{table:sample}, we list the derived properties of the galaxies in our sample, including median $\psi_i$ values and 16\==84\% confidence intervals. 
%An expanded version of the table is available in the electronic copy which further includes \xray\ luminosity for \xray\ detected sources in the 0.5--8 keV band, the $\rm{SFR}$, $M_\star$, source counts, background counts, aperture correction factor and exposure time for the \xray\ data in SB1 and SB2. \\ \indent
%
The mean values of $\psi_i$ and $M_{\star,i}$ versus $t$ are shown in Figure \ref{fig:SFH_all} for three subsamples covering similar redshift ranges. For comparison, we show the SFH for M51 (\citealp{eufrasio,lehmer_17}) and the global SFR density evolution of the Universe (\citealp{madau_14}, \citealp{madau_17}). The SFH for M51 was derived using similar methods to those adopted in this paper, while the global SFH was created using the average SFR density of the Universe normalized to an arbitrary volume to have a similar mass in the oldest age bin to the sample of galaxies in the GOODS fields used throughout this study. \\ \indent
\begin{figure*}
	\centering
	\begin{minipage}{8.75cm}
	\centering
	\includegraphics[width=1\textwidth]{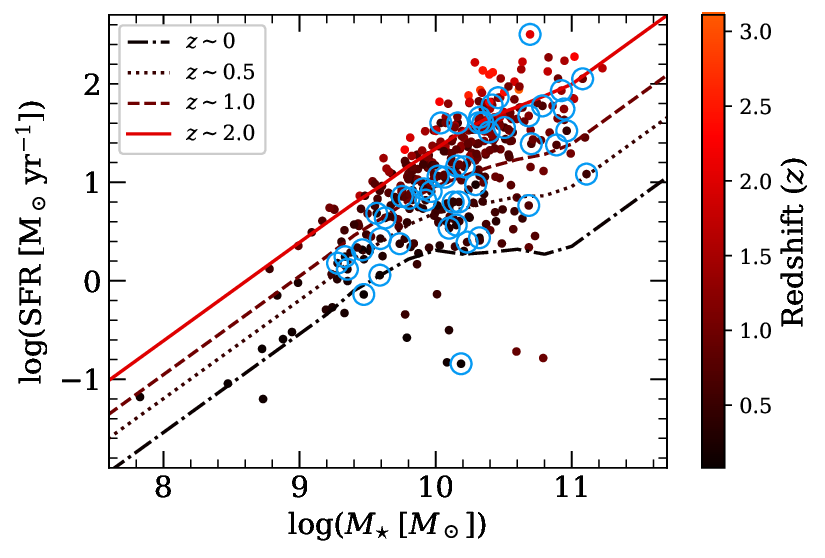}
    \end{minipage} \hfill
    \begin{minipage}{8.75cm}
	\centering
	\includegraphics[width=1\textwidth]{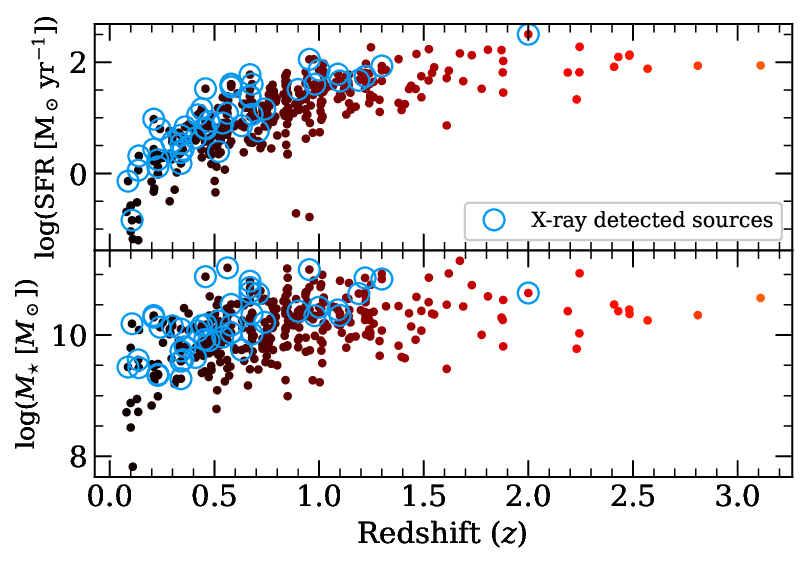}
    \end{minipage}
	
\caption{\textit{Left}: SFR versus stellar mass ($\rm{M}_\star$) for our final sample of 344 galaxies. The four overlaid curves indicate the locations of the galaxy main-sequence for various redshifts that span our sample: $z \sim 0, \; z \sim 0.5, \; z \sim 1.0,\; \text{and} \; z \sim 2.0 $ in the dot-dashed, dotted, dashed, and solid lines, respectively (\citealp{Whitaker_2014}). The data points and main-sequence curves have been colored by redshift for comparison (see color bar). The blue circled points indicate galaxies that were detected in the X-ray (see Section \ref{sub:xray}). \textit{Right}: SFR and stellar mass ($M_\star$) versus redshift for our final sample. Similar to the left, the blue circled points are galaxies that were detected in the X-ray (see Section \ref{sub:xray}), which tend to be relatively nearby and of high SFR and $M_\star$.}
\label{fig:m_sfr_z}
\end{figure*}
\begin{table*} 
\centering 
\begin{footnotesize} 
 \begin{tabular}{ c c c c c c c c c } 
 \hline 
 \hline 
 & & \multicolumn{7}{c}{$\psi \; [M_\odot \: \rm{yr^{-1}}]$} \\ 
\cline{3-9} 
  Source Name (J2000.0) & $z$ & 0--10 Myr & 10--31.6 Myr & 31.6--100 Myr & 0.1--0.316 Gyr & 0.316--1 Gyr & 1--3.16 Gyr & 3.16--10 Gyr \\ [0.5ex] 
  (1) & (2) & (3) & (4) & (5) & (6) & (7) & (8) & (9) \\ 
  \hline 
  \hline 
J033221.43$-$274901.86 & 0.74 & $7.15 \: \pm2.17$ & $4.84 \: \pm3.39$ & $13.13 \: \pm8.51$ & $15.33 \: \pm6.29$ & $4.95 \: \pm3.05$ & $0.82 \: \pm0.59$ & - \\ 
J033219.26$-$274856.17 & 0.68 & $1.65 \: \pm1.11$ & $7.07 \: \pm4.54$ & $14.44 \: \pm8.69$ & $17.39 \: \pm8.09$ & $6.46 \: \pm3.83$ & $3.4 \: \pm1.82$ & - \\ 
J033247.98$-$274855.68 & 0.23 & $0.77 \: \pm0.33$ & $1.4 \: \pm0.94$ & $2.01 \: \pm1.33$ & $1.22 \: \pm0.82$ & $1.48 \: \pm0.64$ & $0.4 \: \pm0.27$ & $0.21 \: \pm0.15$ \\ 
J033241.08$-$274852.97 & 0.34 & $0.46 \: \pm0.3$ & $2.7 \: \pm1.63$ & $4.08 \: \pm2.55$ & $7.41 \: \pm3.69$ & $2.97 \: \pm2.1$ & $8.22 \: \pm1.42$ & - \\ 
J033205.96$-$274845.83 & 0.2 & $0.38 \: \pm0.14$ & $0.52 \: \pm0.35$ & $0.82 \: \pm0.52$ & $0.48 \: \pm0.29$ & $0.2 \: \pm0.13$ & $0.15 \: \pm0.09$ & $0.09 \: \pm0.06$ \\ 
J033224.61$-$274851.52 & 0.58 & $6.54 \: \pm3.47$ & $22.49 \: \pm12.82$ & $50.56 \: \pm25.8$ & $23.66 \: \pm12.41$ & $8.34 \: \pm5.34$ & $3.5 \: \pm2.34$ & - \\ 
J033232.23$-$274845.45 & 0.54 & $2.29 \: \pm1.19$ & $4.95 \: \pm3.5$ & $9.89 \: \pm6.28$ & $16.08 \: \pm7.89$ & $9.25 \: \pm4.53$ & $1.82 \: \pm1.13$ & - \\ 
J033230.55$-$274836.44 & 0.55 & $0.29 \: \pm0.2$ & $0.71 \: \pm0.49$ & $1.44 \: \pm0.94$ & $1.95 \: \pm1.17$ & $1.89 \: \pm0.74$ & $0.43 \: \pm0.26$ & - \\ 
J033253.25$-$274833.50 & 0.23 & $1.64 \: \pm0.23$ & $0.55 \: \pm0.39$ & $0.98 \: \pm0.65$ & $1.19 \: \pm0.65$ & $0.64 \: \pm0.34$ & $0.12 \: \pm0.08$ & $0.09 \: \pm0.06$ \\ 
J033228.48$-$274826.56 & 0.67 & $21.68 \: \pm2.6$ & $2.69 \: \pm2.05$ & $11.37 \: \pm8.33$ & $28.49 \: \pm11.34$ & $13.31 \: \pm7.15$ & $0.88 \: \pm0.66$ & - \\ 
J033234.52$-$274848.50 & 0.21 & $0.68 \: \pm0.42$ & $2.09 \: \pm1.31$ & $3.24 \: \pm1.92$ & $4.73 \: \pm2.91$ & $7.73 \: \pm3.63$ & $5.45 \: \pm2.71$ & $3.06 \: \pm1.58$ \\ 
J033253.09$-$274822.15 & 0.68 & $5.34 \: \pm2.88$ & $9.27 \: \pm6.34$ & $24.26 \: \pm15.25$ & $54.55 \: \pm21.56$ & $21.59 \: \pm11.03$ & $3.48 \: \pm2.35$ & - \\ 
J033251.52$-$274758.05 & 0.74 & $1.98 \: \pm1.33$ & $4.4 \: \pm3.19$ & $8.58 \: \pm5.9$ & $32.43 \: \pm17.14$ & $65.19 \: \pm10.47$ & $4.74 \: \pm3.25$ & - \\ 
J033208.20$-$274752.11 & 0.84 & $30.85 \: \pm6.9$ & $12.3 \: \pm8.13$ & $48.04 \: \pm33.62$ & $67.45 \: \pm24.9$ & $20.43 \: \pm12.5$ & $2.12 \: \pm1.52$ & - \\ 
J033240.06$-$274755.45 & 2.0 & $323.64 \: \pm47.08$ & $250.12 \: \pm131.63$ & $338.31 \: \pm133.86$ & $71.21 \: \pm49.4$ & $30.76 \: \pm22.54$ & - & - \\ 
J033243.26$-$274756.14 & 0.67 & $4.57 \: \pm2.49$ & $9.96 \: \pm6.42$ & $24.19 \: \pm14.29$ & $41.98 \: \pm23.48$ & $46.02 \: \pm14.53$ & $5.42 \: \pm3.72$ & - \\ 
J033205.48$-$274746.90 & 0.9 & $21.39 \: \pm3.26$ & $5.37 \: \pm4.05$ & $16.76 \: \pm10.22$ & $12.45 \: \pm6.37$ & $4.74 \: \pm3.15$ & $0.96 \: \pm0.72$ & - \\ 
J033215.90$-$274750.05 & 0.68 & $4.8 \: \pm2.54$ & $15.24 \: \pm8.62$ & $21.97 \: \pm12.87$ & $12.87 \: \pm7.23$ & $5.3 \: \pm3.29$ & $2.65 \: \pm1.67$ & - \\ 
J033242.29$-$274746.09 & 1.0 & $30.35 \: \pm9.69$ & $32.53 \: \pm20.48$ & $91.08 \: \pm48.55$ & $42.83 \: \pm23.66$ & $22.97 \: \pm13.84$ & $5.57 \: \pm3.72$ & - \\ 
J033237.62$-$274744.07 & 1.1 & $9.6 \: \pm5.0$ & $23.47 \: \pm13.38$ & $58.04 \: \pm30.77$ & $41.48 \: \pm19.48$ & $13.21 \: \pm7.89$ & $4.57 \: \pm3.15$ & - \\ 
  \hline 
\end{tabular} 
\end{footnotesize} 
\caption{Physical properties of our sample. The full version of this table is available electronically and lists the full 344 galaxies and 20 columns. The SFH is displayed as the SFR in each of the seven age-bins in columns 3-9. Galaxies that do not have any star formation in an age-bin due to redshift constraints on age have a dash to represent this lack of contributing stellar mass. Additional columns in the electronic version of the table include: 0.5--8 keV luminosity for X-ray detected galaxies in column 10, $\rm{SFR} \; [\mathit{M_{\odot}} \: \rm{yr^{-1}}]$ (as defined in Section \ref{sub:phys_deriv}) in column 11, total current stellar mass ($M_\star$) in column 12 [$M_{\odot}$]. Columns 13-16 provide observed source counts, background, count estimates, aperture corrections, and exposure times, respectively, for SB1, and columns 17-20 provide the equivalent information for SB2.} 
\label{table:sample} 
\end{table*} 

Figure \ref{fig:SFH_all} shows that, on average, each redshift bin has an elevated SFH compared to the globally averaged SFH of the Universe. It is important to note that $t_{\rm age}$ of the Universe in Figure~\ref{fig:SFH_all} is the cosmic look-back time taken from $z=0$, while each group of galaxies has $t_{\rm age}$ as measured from their rest-frame at $z>0$.  Thus, a value of $t_{\rm age} = 0$ for a galaxy at $z \approx 1$, occurs at a cosmic look-back time of $\approx$7.7~Gyr and will have a peak SFR offset from that of the Universe.  The relatively high SFR values for young populations with ages $<$1~Gyr compared to the older populations is a reflection of our sample being selected for their active star-formation. \\ \indent
The properties of the sample are further illustrated in Figure \ref{fig:m_sfr_z}. On the left we compare our sample SFR and $M_\star$ values to the galaxy main sequence at various redshifts (\citealp{Whitaker_2014}). Although our sample includes a minority of galaxies that fall below the redshift-dependent main sequence, it is clear that the bulk of the sample is near it, or possibly at a slightly elevated SFR. In the right panel, the redshift distribution is shown as it relates to the SFR and $M_\star$ distributions separately. \\ \indent
The sample has a notable dearth of galaxies with SFR~$\simlt 1$~$M_\odot$~yr$^{-1}$ and $M_\star \simlt 10^9 \; M_\odot$, which is a selection effect due to our criteria for sample selection. In particular, only luminous and massive galaxies meet the requirements of having IR detections in six or more bands and high-quality spectroscopic redshifts available. As such, our results are applicable to galaxies above these cutoffs for SFR and $M_\star$, and may not describe well low-mass galaxies or passive elliptical galaxies with low-SFR. Such excluded galaxies may yield different X-ray emission properties to those studied here, as they span broader ranges of metallicity (e.g., low-mass galaxies) and can have hot gas emission associated with deep gravitational potential wells and rich globular cluster LMXB populations (e.g., ellipticals). \\ \indent

\begin{figure}
	\centering
	\includegraphics[width=8.75cm]{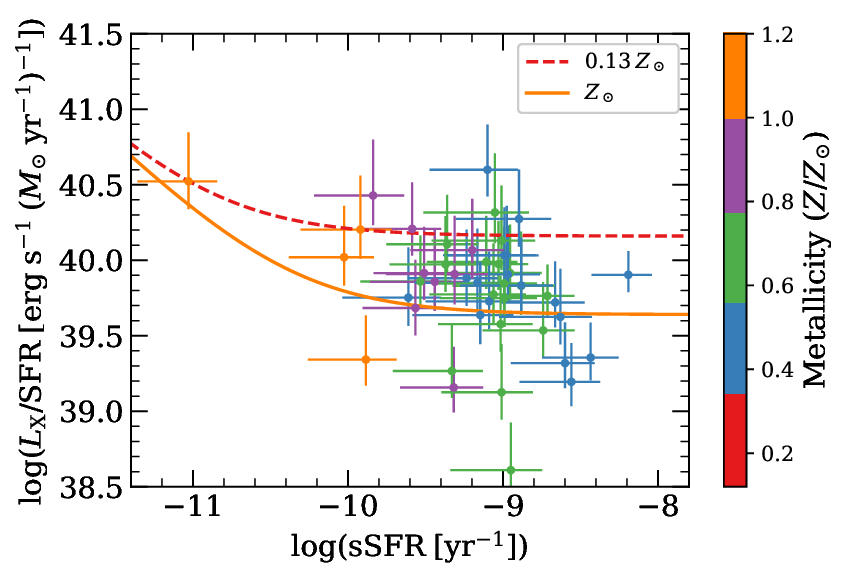}
	\caption{Logarithm of the 2\==10 keV luminosity per SFR (log $L_{\rm X}$/SFR) versus specific SFR (sSFR) for X-ray detected sources in our sample. The X-ray detected sources ({\it circles} with 1$\sigma$ error bars) are compared to local scaling relations, including the $L_{\rm X}/M_\star$ relation for LMXBs from \citet{lehmer_19} and the metallicity-dependent $L_{\rm X}$/SFR relation for HMXBs from \citet{lehmer_21}. Two different relationships are shown corresponding to metallicity values of 0.13 $Z_\odot$ ({\it dashed\/}) and $Z_\odot$ ({\it solid\/}). The detected galaxies largely fall in between the $Z_\odot$ and 0.13 $Z_\odot$ metallicities, which is consistent with the expected metallicity of galaxies in the CDFs.}
	\label{fig:scaling}
\end{figure}
\subsection{X-ray properties of the sample} \label{sub:xray}
After selecting our normal-galaxy sample from the X-ray source catalogs of \cite{xue} and \cite{luo}, we found 10 ($\sim 4 \% $) and 37 ($\sim$20\%) of the galaxies in our sample were X-ray detected in the CDF-N and CDF-S, respectively (see Section \ref{sec:sample_select} for matching procedures). The X-ray detected galaxies in the CDF-N had rest-frame 2\==10~keV luminosities in the range of $4.1 \times 10^{39} \; \rm{erg \; s^{-1}}$ to $1.1 \times 10^{42} \; \rm{erg \; s^{-1}}$, while in the CDF-S, the galaxy luminosities ranged from $6.0 \times 10^{39} \; \rm{erg \; s^{-1}}$ to $2.5 \times 10^{42} \; \rm{erg \; s^{-1}}$. The physical properties of these \xray\ detected galaxies are also known, as derived from the SED fitting procedures outlined in Section \ref{sub:phys_deriv}. These \xray\ detected galaxies are highlighted with blue circles in Figure \ref{fig:m_sfr_z}. The stellar masses span the range of $10^{9.3}$\==$10^{11.1} \; M_\odot$ and the SFRs span 0.2\==$420 \; M_\odot \; \rm{yr}^{-1}$. For a given redshift, the \xray\ detected galaxies tend to be the most massive and highest-SFR galaxies in the sample. \\ \indent
In Figure~\ref{fig:scaling}, we show the 2\==10~keV luminosity per unit SFR ($L_{\rm X}$/$\rm{SFR}$) versus sSFR for the 47 \xray\ detected galaxies in our sample, accounting for uncertainties in $L_{\rm X}$, SFR, and $M_\star$. For comparison, we overlaid the $L_{\rm{X}} / \rm{SFR} \; \rm{vs} \; \rm{sSFR}$ scaling relations from \cite{lehmer_19, lehmer_21}. \cite{lehmer_19} provides a scaling relation for $\alpha_{\rm{LMXB}}$ ($L_{\rm X}$[LMXB]/$M_\star$) in late-type galaxies, and \cite{lehmer_21} provides a metallicity-dependent scaling of $\beta_{\rm HMXB}$ ($L_{\rm X}$[HMXB]/SFR). Figure \ref{fig:scaling} shows the predicted relations based on the single $\alpha_{\rm{LMXB}}$ value and $\beta_{\rm HMXB}$ at two different metallicities ($Z \sim$~0.13 and 1~$Z_\odot$). The single $\alpha_{\rm{LMXB}}$ value determines the slope of the line as it enters from the left, while each of the $\beta_{\rm HMXB}$ values set the baseline for each line. The galaxies in the CDFs are expected to have somewhat lower metallicities than typical galaxies in the local universe, due to less chemical enrichment having taken place at higher redshifts. In order to verify this we estimate the metallicities using the fundamental metallicity relation (Equation 2 from \citealp{mannucci_10}). Mannucci metalicity values are translated to solar units using consistent methodology to previous comparison studies to avoid introducing large systematic errors (\citealp{kewley_08}). The resulting metallicities have a range of 0.40\==1.16~$Z_{\odot}$ with a mean of $0.68 \; Z_{\odot}$. While this does not reach levels of $0.13 \; Z_{\odot}$ as shown in Figure \ref{fig:scaling}, these do reflect the expected lower metallicity values, resulting in elevated scaling relations. \\ \indent
\begin{figure}
	\centering
	\includegraphics[width=8.75cm]{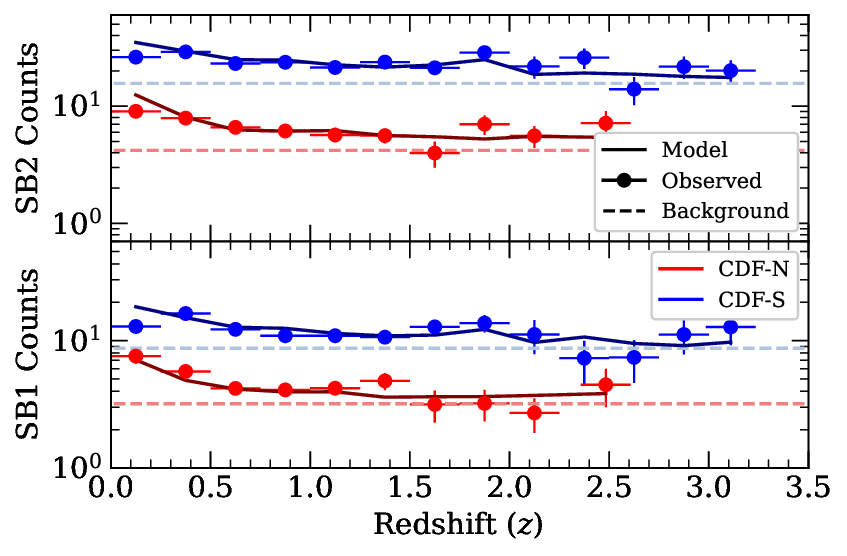}
	\caption{The source, average background, and model counts for galaxy subsets in this study shown in points, light dashed lines, and dark solid lines respectively. Blue (red) filled circles and lines represent results from the CDF-S (CDF-N). The top and bottom panels show SB1 (0.5\==1 keV) and SB2 (1\==2 keV) results, respectively. Galaxies were grouped into subsets based on redshift to clearly show the entire data set without overcrowding. Each displayed data point represents the mean number of counts per source (and 1$\sigma$ error) in the subset, which have redshift selection bins indicated by horizontal error bars. The displayed model gives the bin-by-bin predicted counts from our fully independent model described in Sections~\ref{sub:mod} and \ref{sub:age_dependence}.}
	\label{fig:xray_counts}
\end{figure}
The majority of the galaxies in our sample (297 of 344; $\sim$86\%) are undetected in the CDFs. To incorporate the \xray\ constraints from these galaxies, we extracted the \xray\ counts centered on the optical positions of all galaxies in our sample. In this process, we extracted counts from each galaxy using either a 2.5~arcsec or 1.5~arcsec radius circular aperture, depending on whether the source was at $z < 0.7$ or $z \ge 0.7$, respectively. These apertures encircle rest-frame radii of $\sim$10 kpc for all galaxies in our sample, a size that was adopted to both ensure extraction of the majority of the emission from the galaxies while maintaining high signal-to-noise ratio for the galaxies. 
Background count contributions to the extraction regions were estimated using background and exposure maps from \citet{xue} and \citet{luo}. Here, background counts and cumulative exposure-map values were first extracted from the maps using relatively large, 25~arcsec radii apertures centered on the sources to obtain well-constrained estimates of the local background rates (counts per second per pixel). We expect that local background measurements are good measurements of the mean of the local Poisson distributions. Then, for a given source, the background rate was re-scaled to the cumulative exposure-map values obtained from within the source aperture to estimate model background-count contributions to the galaxy count extractions. \\ \indent
We chose to extract \xray\ counts (using the above procedure) using two \xray\ bands: the 0.5\==1~keV (SB1) and 1\==2~keV (SB2) bands. These bands were selected to optimize \textit{Chandra}'s response and sensitivity to components dominating normal-galaxy \xray\ emission (hot gas and XRBs). The use of multiple bands allows for constraints on spectral shape, which is sensitive to the ratio of XRB and hot gas emission (see Section \ref{sub:mod} and Figure \ref{fig:omega_variance}). We chose to exclude the 2\==8~keV (HB) bandpass from our analyses, since the stacked HB emission from similar galaxy samples has been shown to have non-negligible signatures from heavily-obscured and Compton-thick AGN \citep[e.g.,][]{xue2012}. The SB emission can be sensitive to weaker AGN when using stacking procedures, however using this method (along with the conservative sample selection process outlined in Section \ref{sec:sample_select}) we do not expect weak AGN contamination to significantly affect our results.\\ \indent
%energy ranges were left out in the case of any remaining potential AGN contamination. If an AGN remained in the sample it would have a much higher HB emission than a normal galaxy, allowing it to significantly swing the spectral shape of the model, which will be outlined in Section \ref{sec:tech}.
%
For illustrative purposes, we created Figure \ref{fig:xray_counts}, which shows the raw extracted SB1 and SB2 source counts for galaxy subsamples in the CDF-N and CDF-S, including both \xray\ detected and non-\xray\ detected sources. The galaxy subsamples were constructed by grouping sources in bins of redshift; the data points indicate mean counts per galaxy for each grouping. For comparison, the mean background levels for the full CDF-N and CDF-S samples are shown as horizontal dashed lines. From Figure \ref{fig:xray_counts}, it is apparent that the \xray\ counts from galaxy populations across the full redshift range provide some signal above the nominal background level. The exception is that of the SB1 constraints for galaxy populations at $z \simgt$~1\==2, where the constraints are largely consistent with the background level. We note here that the grouping of subsamples is for illustrative purposes only, and these groupings are not used in the analysis throughout this paper. In Section~\ref{sec:tech}, we utilize raw constraints for galaxies on an individual basis, in the form of extracted SB1 and SB2 counts at the source positions, along with the SFH information obtained in Section~\ref{sub:phys_deriv} to construct a self-consistent model for how the \xray\ SED from galaxies evolves as a function of age. \\ \indent
%0.5 arcsecond apertures centered on the optical locations of the galaxies in our sample, along with local background estimates. The data was put into randomized groups of 10 galaxies, with the exception of the final remainder galaxies in each field. The black points in each section show the median signal counts of the selected galaxies, while the gray points show the median background counts, and the vertical lines show the standard deviation for each group. The figure also shows the difference between the CDF-S and CDF-N left to right, as well as SB1 and SB2 bottom to top. The \xray\ counts in these bands will the data that we use in our modeling of $L_{\rm{X}} (\rm{t})$\\ \indent
%
\section{Modeling Techniques} \label{sec:tech}
\subsection{Model Construction} \label{sub:mod}
To self-consistently model the \xray\ counts from our full galaxy sample, we made use of a forward modeling approach that incorporates both local estimates of the background and a SFH-dependent model of the \xray\ emission from galaxies. For a given bandpass with energy boundaries $E_1$\==$E_2$, the observed counts from each source, $S^{\rm obs}_{E_1-E_2}$, can be modeled as follows:
\begin{equation}
S^{\rm model}_{E_1-E_2} = \frac{\Phi_{E_1-E_2} \times t_{E_1-E_2}}{\xi_{E_1-E_2}} + B_{E_1-E_2},
\end{equation}
where $\Phi_{E_1-E_2}$ is the model count-rate for the source, which will depend on both physical properties, including galaxy redshift, SFH, and SED shape (see below for details), and the instrumental response, as per the energy-dependent auxiliary response file (ARF). $t_{E_1-E_2}$ is the vignetting-corrected exposure time at the location of the galaxy, $\xi_{E_1-E_2}$ is the aperture correction, which adjusts the total model count-rate to the count rate expected within the extraction aperture (typically just above 1 for our sample), and $B_{E_1-E_2}$ is the local background estimate (see Section~\ref{sub:xray}).

The model count rate for a given bandpass can be calculated as follows:
\begin{equation}
\Phi_{E_1-E_2} = k_{E_1-E_2} F_{E_1-E_2} = \frac{k_{E_1-E_2}}{4 \pi d_L^2} \int_{E_1(1+z)}^{E_2(1+z)} \ell(E^\prime) \; dE^\prime,
\end{equation}
where $k_{E_1-E_2}$ is a conversion factor that converts flux to count-rate (counts~s$^{-1}$), given an \xray\ SED shape, $F_{E_1-E_2}$ is the bandpass integrated flux (ergs~cm$^{-2}$~s$^{-1}$, $d_L$ is the luminosity distance (cm), and $\ell(E^\prime)$ is the rest-frame X-ray SED model, in units of ergs~s~keV$^{-1}$ (see below). Since our goal is to obtain a model of the rest-frame \xray\ SED shape, $k_{E_1-E_2}$ is model dependent, and can be calculated as
\begin{equation}
k_{E_1-E_2} = \frac{\int_{E_1}^{E_2} (1+z) \; \ell\,[E(1+z)] \; dE}{\int_{E_1}^{E_2} (1+z) \; \ell\,[E(1+z)] \; {\rm ARF}(E)/E \; dE},
\end{equation}
\\
where ARF$(E)$ is the energy-dependent effective area curve appropriate for the central regions of each of the CDF fields. The off-axis variation of the ARF is accounted for in the vignetting-corrected exposure time.

Our intrinsic SFH\==and\==SED-dependent model, $\ell(E)$, is modeled for a given source as the linear combination of contributions to the overall source spectrum from each of the seven SFH age bins (see Section \ref{sub:phys_deriv}). For the $i$th age bin, the contributing \xray\ SED is modeled as:
\begin{equation}
\ell_i(E) = \frac{\gamma_i M_{\star,i} \left[ f_{{\rm gas}, i}(E) + f_{{\rm XRB}, i}(E) \right]}{\int_{\rm 2~keV}^{\rm 10~keV} \left(f_{{\rm gas}, i}[E] + f_{{\rm XRB}, i}[E]\right) dE}, 
\end{equation}
\\
where $\gamma_i$ is a free parameter that is equivalent to the 2\==10~keV luminosity per unit stellar mass, $L_{\rm X}/M_\star$, contribution associated with the stellar population in the $i$th age bin, and $M_{\star,i}$ is the contribution to the total stellar mass of the galaxy from the $i$th age bin (see lower-right panel of Fig.~\ref{fig:SED} for an example of the seven values of $M_{\star,i}$ for a galaxy). The \xray\ SED shape in the $i$th bin is modeled by the arbitrarily normalized $f_{{\rm gas},i}$ and $f_{{\rm XRB},i}$ terms, which consist of fixed SED shapes that are characteristic of hot gas and XRBs, respectively, in the local universe. Specifically, we used an {\ttfamily xspec} parameterization of the average SED shapes for hot gas and XRBs, as determined for spatially-resolved \xray\ studies of M83, NGC 253, NGC 3256, and NGC 3310 using the combination of \chandra\ or \xmm\ and \nustar\ data across the 0.3--30~keV spectral range \citep[see][]{Wik_2014,lehmer_15,Yukita_2016}.
We approximated the average hot gas SED shape as an absorbed ($N_{\rm H, gas} =  5.5 \times 10^{21}$~cm$^{-2}$) single-temperature thermal plasma ({\ttfamily apec}) with $kT =$~0.2~keV, and the XRB spectrum was modeled as an absorbed ($N_{\rm H, XRB} = 3 \times 10^{21}$~cm$^{-2}$) broken power-law ({\ttfamily bknpow}) with $\Gamma_1 = 1.8$, $E_{\rm break} = 5$~keV, and $\Gamma_2 = 2.6$.

Although the SED shapes $f_{{\rm gas},i}$ and $f_{{\rm XRB},i}$ are fixed, the relative contribution of XRBs to the total \xray\ luminosity is modeled as a single free parameter (per age bin) as:
\begin{equation}
\omega_i \equiv \frac{\int_{\rm 0.5~keV}^{\rm 8~keV} f_{{\rm XRB},i}(E) \; dE}{\int_{\rm 0.5~keV}^{\rm 8~keV} \left(f_{{\rm gas}, i}[E] + f_{{\rm XRB},i}[E] \right)  \; dE}.
\end{equation}
\\
\begin{figure}
	\centering
	\includegraphics[width=8.75cm]{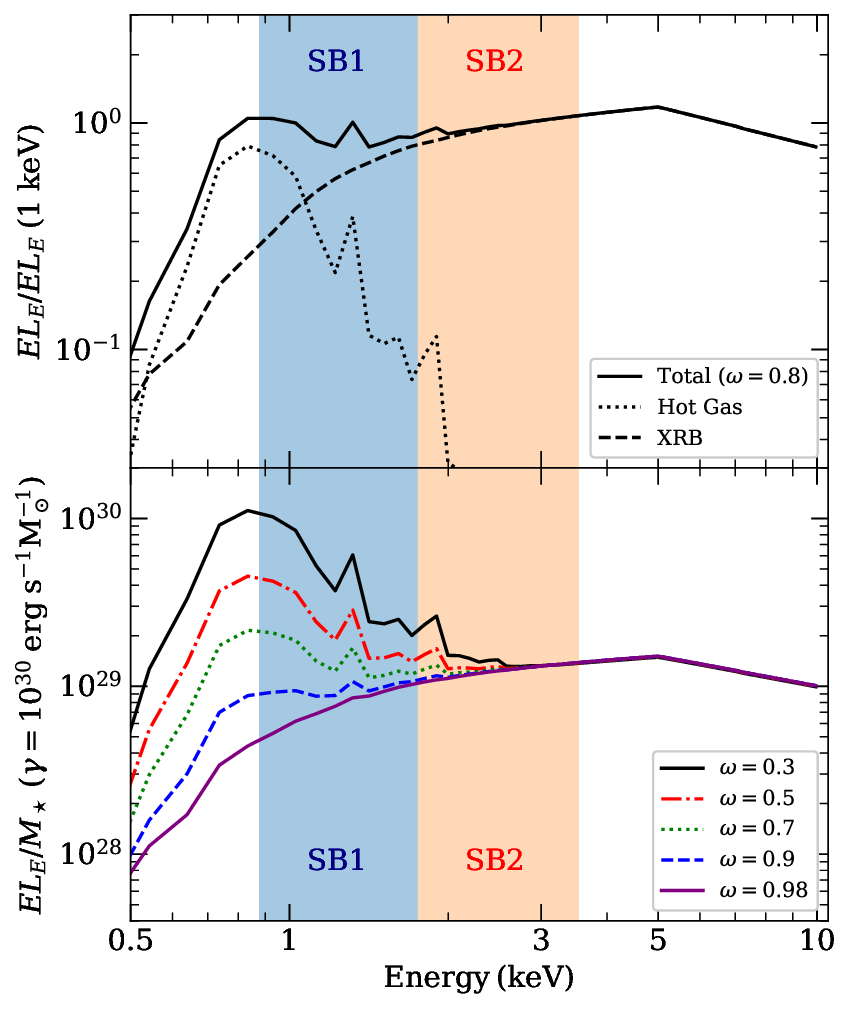}
	\caption{\textit{Top}: A decomposition of the spectral model used in this study, showing the hot gas and XRB model components as dotted and dashed lines respectively, with the total model spectrum shown as a solid line. The displayed spectrum demonstrates the spectral shape of $\omega=0.8$, similar to the value found for the semi-independent model in Section \ref{sub:fit}. \textit{Bottom}: The variation of model SED shape as a function of $\omega$ (the fraction of the total 0.5\==8 keV luminosity from the XRB model component). These spectra adopt a fixed value of $\gamma = 10^{30}$~erg~s$^{-1}$~$M_\odot^{-1}$, the 2\==10 keV luminosity per stellar mass, and therefore primarily show variations in the hot gas contribution. In both top and bottom panels, the blue and orange shaded regions show the SB1 and SB2 rest-frame energy ranges for $z=0.76$, the median redshift of our sample.}
	\label{fig:omega_variance}
\end{figure}
In Figure \ref{fig:omega_variance}, we show how the value of $\omega$ impacts the shape of the model \xray\ SED. The value of $\omega$ is bound between 0 and 1 as required by the definition, although any value $<0.5$ implies that the hot gas component is more luminous than the XRB population. Additionally, it is important to recall that the total spectrum is normalized in the 2\==10 keV band via $\gamma$. Our choice to define $\omega$ in terms of the 0.5\==8~keV band, instead of the 2\==10~keV band (as used for $\gamma$), was motivated by the fact that the hot gas emission drops off precipitously above $\approx$1.5~keV leaving the XRB emission dominant at higher energies by several orders of magnitude. \\ \indent
To summarize, our overall model consists of the free parameters $\gamma_i$ and $\omega_i$, which specify the normalization and spectral shape, respectively, at each of the seven age bins. To constrain these parameters, we made use of the aperture-extracted counts for all sources in both SB1 and SB2 bandpasses (Col. 13 and 17 in Table \ref{table:sample}), along with the count predictions from our model, using the free parameters and the SED-derived masses (see Eqn.~(4)); we describe our statistical analyses in the next section. We explore two model scenarios: (1) a {\it fully independent model} in which all values of $\gamma_i$ (or $L_{{\rm X}}/M_\star$ for the $i$th age bin) and $\omega_i$ (spectral shape) were treated as independent fitting parameters (14 total parameters); and (2) a {\it semi-independent model} where we fit for seven values of $\gamma_i$, but identify a single best spectral shape where $\omega_i = \omega_{\rm best}$ applied to all age bins (8 total parameters). \\ 
\subsection{Model Fitting and Parameter Estimation} \label{sub:fit}
\begin{figure}
	\centering
	\includegraphics[width=8.75cm]{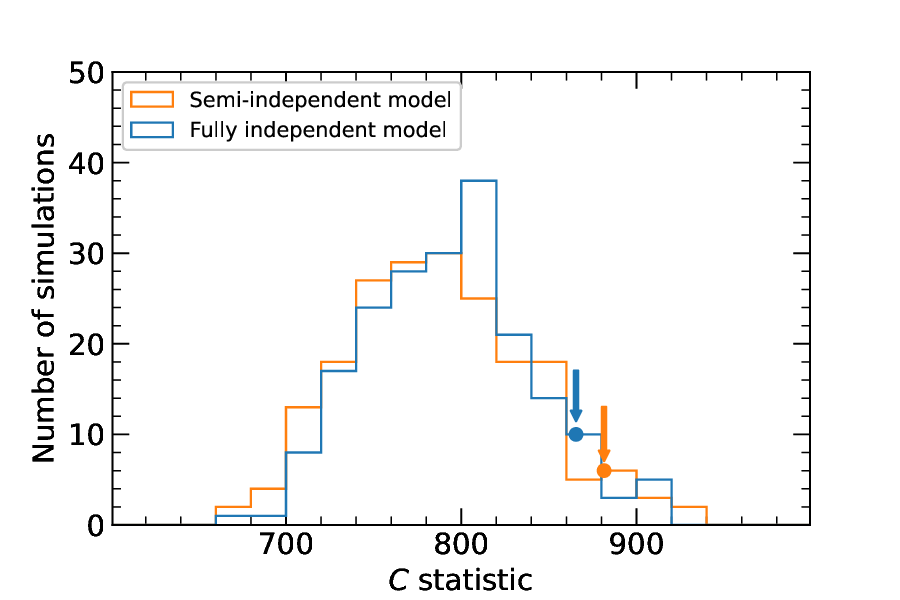}
	\caption{The expected $C$ statistic distributions for our fully independent and semi-independent model fits (blue and orange, respectively; see Section \ref{sub:mod}). These distributions were created using 200 simulated data sets drawn from the best-fit model (see Section \ref{sub:fit}). The points and downward-pointing arrows show the location of the values of $C$ found for each model fit to the actual data. The numerical values of $C$ and null-hypothesis probabilities are provided in Table \ref{table:parameters}.}
	\label{fig:c_dist}
\end{figure}

All statistical calculations were performed using total observed counts $S_{E_1-E_2}^{\rm obs}$ at each source location, for both the SB1 and SB2 bands, and the corresponding model counts, $S_{E_1-E_2}^{\rm model}$, as calculated following Eqn.~(1)\==(5). This approach allows us to simultaneously use the statistics of each and every source, and differs from conventional stacking approaches \citep[e.g.,][]{Basu_Zych_2013b,lehmer_16,fornasini_19,fornasini_20}, which effectively reduce the available degrees of freedom by considering a single "stacked" count-rate for a galaxy sample. \\ \indent
To fit model parameters and sample the posterior distribution function, we made use of the $C$ statistic (e.g., \citealp{cash, Kaastra_2017}), which is calculated as
\begin{equation}
C = 2 \sum_{j=1}^{n} \sum_{k=1}^2 S_{j,k}^{\rm model} - S_{j,k}^{\rm obs} + S_{j,k}^{\rm obs} \ln(S_{j,k}^{\rm obs}/S_{j,k}^{\rm model}).
\end{equation}
Here, the indices $j$ and $k$ correspond to the $n$ galaxies and two bandpasses (i.e., $k = 1$ and 2 respectively represent SB1 and SB2 bands), respectively. As such, the summations contain a total of 688 independent terms (i.e. $n=344$ galaxies and two bands). 

For a given model scenario, we identified best-fit parameters by minimizing $C$ in Eqn.~(6), and we sampled their posterior distributions using a Monte Carlo (MC) approach. In our procedure, we used the dual annealing optimization method \citep[][]{xiang_97} provided in the {\ttfamily Python}-based {\ttfamily SciPy} package\footnote{https://www.scipy.org/} to find the minimum $C$ for a given model. We define the ``best-fit'' solution as the set of parameter values that minimize $C$ when adopting the best-fit age-dependent masses from our SED fitting results described in Section~\ref{sub:phys_deriv}.
%while using the measured number of \xray\ counts for each source as well as the age-binned masses ($M_{\star , i}$) from the SED fitting procedure which resulted in the lowest $\chi^2$ value.\\ \indent

\begin{figure*}[t]
	\centering
	\includegraphics[width=\textwidth]{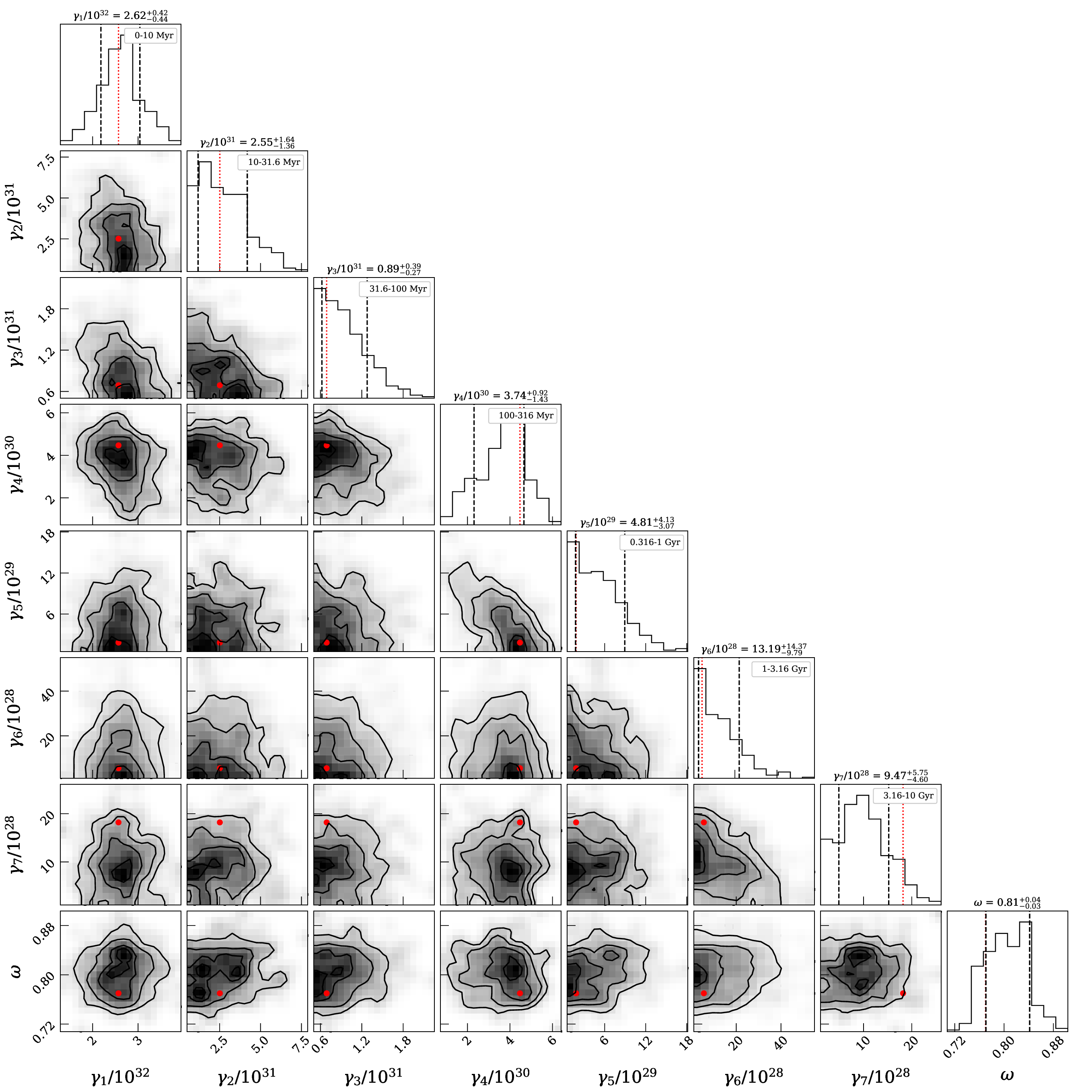}
	\caption{Marginalized probability distributions for the 8-parameters in our semi-independent model. 1D distributions are shown along the diagonal and 2D distributions are shown below the diagonal. In the 1D plots, the vertical dashed black lines indicate 16\==84\% confidence ranges; we also provide annotations listing the median and 16\==84\% confidence ranges of the parameters ({\it above each plot\/}) and the age bins associated with each of the $\gamma$ parameters ({\it upper-right corner of each plot\/}). The plots below the diagonal show the 2D marginalized distributions, and use contours and shading to show probability density. The red points on the 2D plots, and red dotted lines on the 1D plots, show the single best-fit values, as defined by the minimum $C$ value when using ideal masses and counts. Each of the parameters are shown in linear space, and for clarity, $\gamma$ values are normalized to appropriate scales, as indicated in the axis labels and annotations. All values of $\gamma \; (L_{\rm X} / M_\star)$ are provided in units of erg~s$^{-1}$~$M_\odot^{-1}$.}
	\label{fig:param_dist}
\end{figure*}

\begin{figure*}[t]
	\centering
	\begin{minipage}{8.75cm}
	\centering
	\includegraphics[width=\textwidth]{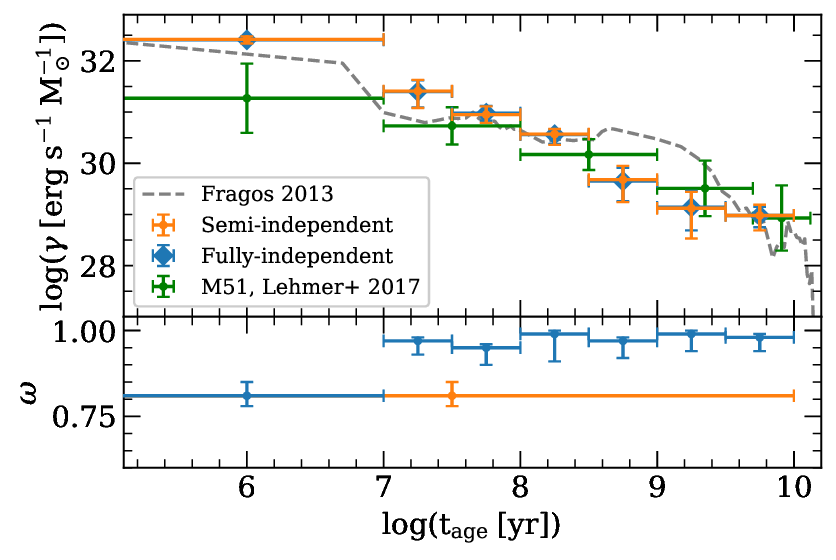}
	\end{minipage} \hfill
	\begin{minipage}{8.75cm}
	\centering
	\includegraphics[width=\textwidth]{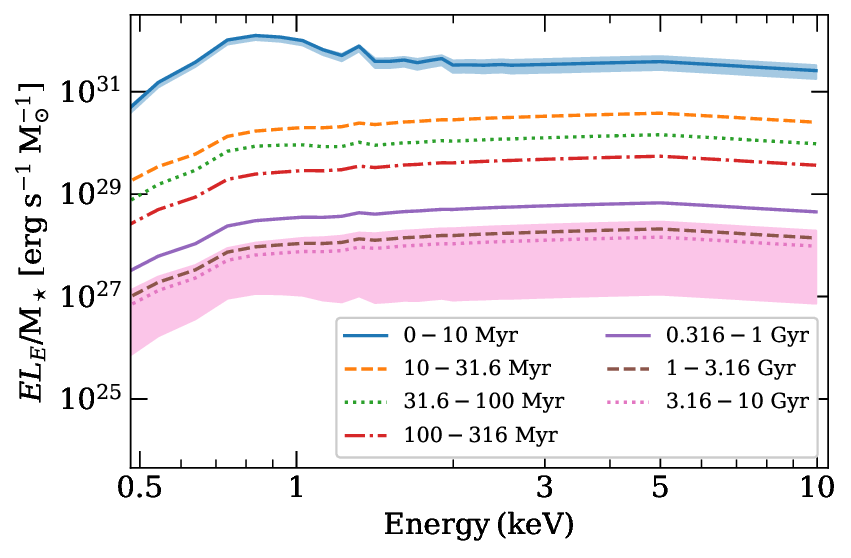}
	\end{minipage}
	\caption{
	%%%
	\textit{Top Left}: The age-evolution of $\gamma_i$ in the 2\==10 keV band. The horizontal error bars show the width of each age-bin, while the vertical error bars indicate the 16$-$84\% confidence ranges shown in Table \ref{table:parameters}. The blue diamonds of the fully independent model are slightly enlarged for visibility as they largely overlap with the orange points of the semi-independent model. The results from this study are compared to the theoretical results from \citet{fragos_13a} (shown as a gray dashed line) and a similar study performed on M51 in \citet{lehmer_17} (shown in green points). \textit{Bottom Left}: The age-evolution of $\omega$ for each of the two models, using the same colors as above. The horizontal and vertical error bars represent the same quantities as above as well.
	%%%
	\textit{Right}: Resulting stellar-mass normalized X-ray SED models, $E L_E/M_\star$, for the seven age-bins, based on our fully independent model (i.e., including variations in both $\gamma$ and $\omega$ with age). The displayed SEDs correspond to those produced by the median parameter values listed in Table \ref{table:parameters}; various colors and linestyles correspond to age bins as annotated in the legend. To give a sense of uncertainties, we show shaded regions that represent the full span of models with the lowest 68\% of $C$ values (see Section \ref{sub:fit}) for the youngest and oldest age bins.
	%%%
	}
	\label{fig:LxM}
\end{figure*}

To sample the posterior distribution for our model parameter sets, we repeated the dual annealing optimization for perturbed values of the age-binned masses and X-ray counts. The age-binned masses were perturbed by pulling values of $M_{\star , i}$ in accordance with their posterior distributions (see Section \ref{sub:phys_deriv}). The \xray\ count distributions were perturbed assuming that they follow Poisson distributions with mean values equal to the measured counts, signal plus background, for each source. We repeated the process of pulling perturbed sets of age-binned masses and counts for our sample 1000 times and determined optimized parameter sets for each perturbed set using the dual annealing method. The list of resulting parameters provides a sampling of the posterior distribution, and the list of a given parameter provides a distribution of its 1D marginalized distribution function; we used these distributions to calculate confidence intervals for the parameters.\\ \indent
To test whether a given best-fit model was statistically acceptable (i.e., calculate a "goodness of fit" to determine whether the data were consistent with the null hypothesis), we performed simulations, where simulated data and distributions of $C$ were generated from the best-fit model, and then refit to determine expected values of $C$. For a given simulation, we (1) chose values of $M_{\star, i}$ from the SED-fitting-based posteriors from {\ttfamily Lightning}; (2) used our best-fit model with the selected $M_{\star, i}$ values to estimate model \xray\ luminosities for each of the galaxies; (3) perturbed those luminosities to simulate scatter in the relations, following the SFR and $M_\star$ dependent scatter values given in Section~5.3 of \citet{lehmer_19}; (4) converted these perturbed luminosities into predicted source counts following equations (1)\==(4) using our best-fit model parameters; (5) perturbed the predicted source counts (assuming Poisson distributions) to obtain simulated counts; and finally, (6) re-fit our model to the simulated counts to identify the minimum $C$ for the simulation. This procedure was repeated 200 times for each model, providing a distribution of expected values of $C$ and a means for comparing the minimum $C$ value obtained from our data with that expected from the best-fit model. \\ \indent
\begin{table*} 
\begin{center}
 \begin{tabular}{lcccccc}
 \hline
 \hline
 &  & \multicolumn{2}{c}{\textbf{fully independent model}} & & \multicolumn{2}{c}{\textbf{semi-independent model}} \\
  \cline{3-4}
  \cline{6-7} 
 Age Range & Bin Number & $\rm{log}(\gamma \; [erg \; s^{-1} \; M_\odot^{-1}])$ & $\omega$ & & $\rm{log}(\gamma \; [erg \; s^{-1} \; M_\odot^{-1}])$ & $\omega$ \\ 
  \hline
  \hline
 0--10 Myr & 1 & $32.41 \; (32.41^{+0.07}_{-0.06})$ & $0.75 \; (0.81^{+0.04}_{-0.03})$ & & $32.41 \; (32.42^{+0.06}_{-0.08})$ & $0.77 \; (0.81^{+0.04}_{-0.03})$ \\
 10--31.6 Myr & 2 & $31.36 \; (31.40^{+0.22}_{-0.31})$ & $0.97 \; (0.98^{+0.01}_{-0.04})$ & & $31.40 (31.41^{+0.22}_{-0.33})$ & \vdots \\
 31.6--100 Myr & 3 & $31.00 \; (30.98^{+0.14}_{-0.15})$ & $0.95 \; (0.97^{+0.01}_{-0.05})$ & & $30.84 \; (30.95^{+0.16}_{-0.16})$ & \vdots \\
 100--316 Myr & 4 & $30.70 \; (30.56^{+0.10}_{-0.18})$ & $0.99 \; (0.98^{+0.01}_{-0.08})$ & & $30.65 \; (30.57^{+0.10}_{-0.21})$ & \vdots \\
 0.316--1 Gyr & 5 & $29.42 \; (29.65^{+0.26}_{-0.39})$ & $0.97 \; (0.98^{+0.01}_{-0.05})$ & & $29.26 \; (29.68^{+0.27}_{-0.44})$ & \vdots \\
 1--3.16 Gyr & 6 & $28.36 \; (29.14^{+0.32}_{-0.45})$ & $0.99 \; (0.98^{+0.01}_{-0.05})$ & & $28.74 \; (29.12^{+0.32}_{-0.59})$ & \vdots \\
 3.16--10 Gyr & 7 & $29.15 \; (28.98^{+0.17}_{-0.23})$ & $0.98 \; (0.98^{+0.01}_{-0.04})$ & & $29.26 \; (28.98^{+0.21}_{-0.29})$ & \vdots \\ 
 \hline
 \hline
 \multicolumn{7}{c}{$C$ statistic analysis} \\
 \hline
 \hline
 $C_{\rm{min}}$ & & \multicolumn{2}{c}{866} & & \multicolumn{2}{c}{882} \\
 (SB1, SB2) & & \multicolumn{2}{c}{(429, 437)} & & \multicolumn{2}{c}{(440, 442)} \\
 (CDF-S, CDF-N) & & \multicolumn{2}{c}{(414, 452)} & & \multicolumn{2}{c}{(418, 464)} \\
 $P_{\rm{null}}$ & &\multicolumn{2}{c}{0.080} & & \multicolumn{2}{c}{0.045} \\
 \hline
\end{tabular}
\end{center}
\caption{Parameter distributions and $C$ statistics for the two models. The leftmost column in the upper portion shows the age ranges for each of the seven age bins, with the following four columns showing the $\gamma_{\rm{i}}$ and $\omega_{\rm{i}}$ values for the two models. The semi-independent model uses the same $\omega$ value for all seven age bins. Two values are displayed in each of the columns: the best-fit value and the median with 16--84\% confidence ranges. The best-fit values show the values when using the ideal masses and counts (which result in the $C$ values in the lower portion of the table), while the median and confidence ranges come from the 1D marginalized distributions created from the MC method (see section \ref{sub:fit}). In the lower portion, the calculated minimum $C_{\rm{min}}$ is listed for each model and the $P_{\rm null}$. The $C_{\rm{min}}$ is displayed as a total first, and then broken down into contributions from SB1 and SB2 on the following lines, and then contributions from CDF-S and CDF-N in the third line. The $P_{\rm null}$ value is calculated by simulating fake data sets around the model and then refitting those data sets (see Section \ref{sub:fit}), reflecting the probability of the model properly fitting the data.\label{table:parameters}}
\end{table*}
The distributions of the $C$ statistic for our fully independent and semi-independent models are shown, along with the values of the minimum $C$ obtained for the models, in Figure \ref{fig:c_dist}. The $P_{\rm null}$ value (probability that the data are consistent with being drawn from the model) is calculated as the number of the simulated trials that resulted in a $C$ value greater than that obtained for our best-fit model, $N(C>C_{\rm min})$, divided by the total number of simulated trials (i.e., $N_{\rm total}=200$): 
\begin{equation}
P_{\rm null} = \frac{N(C>C_{\rm min})}{N_{\rm total}}
\end{equation}
%
%This functions as a one-way, discrete approximation to what would normally be a z-test for a Gaussian distribution. 
We consider $P_{\rm null}$ value less than 0.01 (or greater than 0.99) as unacceptable in terms of the data being consistent with the model. We find that both models are acceptable ($P_{\rm null} > 0.01$), with the fully independent model having a somewhat higher (but not significantly higher) value of $P_{\rm null}$ than the semi-independent model. Both models producing low values for $P_{\rm null}$ indicates that this model does simplify the relation between age and these parameters ($\gamma$ and $\omega$), but it is not an over simplification to the point of statistical significance. \\ \indent
In Table \ref{table:parameters}, we summarize the results of the fully independent and semi-independent model fits, including parameter estimations and statistical evaluations of our models ($C$ and $P_{\rm null}$). In Figure~\ref{fig:xray_counts}, we overlay semi-independent model-predicted raw numbers of counts as a function of redshift in both bandpasses (i.e., SB1 and SB2) and both survey fields (CDF-N and CDF-S). The lack of any obvious outliers in this representation indicates that our model provides a reasonable description of the galaxy counts in both bandpasses, for sources in both observational fields and across the full redshift range of our sample. In the lower portion of Table~\ref{table:parameters} we show the total $C_{\rm min}$ as well as the contributions from SB1 and SB2 separately, followed by the contributions from the CDF-S and CDF-N. The models do not appear to significantly favor any one split over the other in any way. In Section \ref{sub:subs}, we fit the CDF-S and CDF-N separately, and further assess the consistency between results. \\ \indent
In Table~\ref{table:parameters}, we also list the resulting fit parameters, $\gamma$ and $\omega$, starting with their best-fit values (as defined above), followed by their medians and 16\==84\% confidence ranges, which were derived from our MC 1D marginalized parameter distributions. Values of the $C$ statistic are shown for the best-fit models. \\ \indent
For illustrative purposes, we created Figure~\ref{fig:param_dist}, which shows the 1D and 2D marginalized distributions for the semi-independent model. It is clear that some of the parameters are correlated (e.g., $\gamma_1$ versus $\gamma_2$), and others show distributions consistent with being upper limits (e.g., $\gamma_2$, $\gamma_5$, and $\gamma_6$). Nonetheless, the full solution is informative, in terms of constraining the age-dependent evolution of \xray\ emission. \\ \indent
In the left panels of Figure~\ref{fig:LxM}, we graphically show our resulting age-dependence of $\gamma$ and $\omega$ for both the fully independent and semi-independent models, and in the right panel of Figure~\ref{fig:LxM}, we display the resulting stellar-mass normalized SED at all seven age bins for the fully independent model. Despite the allowed evolution of spectral shape for the fully independent model, both models provide very similar predictions of $\gamma$ as a function of age, showing a marked decline of $\approx$3 orders of magnitude from 10~Myr to 10~Gyr. The fully independent model fit prefers evolution of $\omega$ with stellar age, such that the \xray\ SED transitions from requiring a significant hot gas component for the youngest age bin to preferring a more XRB dominant SED for all subsequent age bins. This results in a hardening of the \xray\ SED as stellar age increases. In the next section, we provide physical interpretation of these results and discuss them in the context of previous studies. \\ \indent
\section{Discussion} \label{sec:discussion}
\subsection{The Stellar-Age Dependence of X-ray Emission} \label{sub:age_dependence}
\begin{figure}
	\centering
	\includegraphics[width=8.75cm]{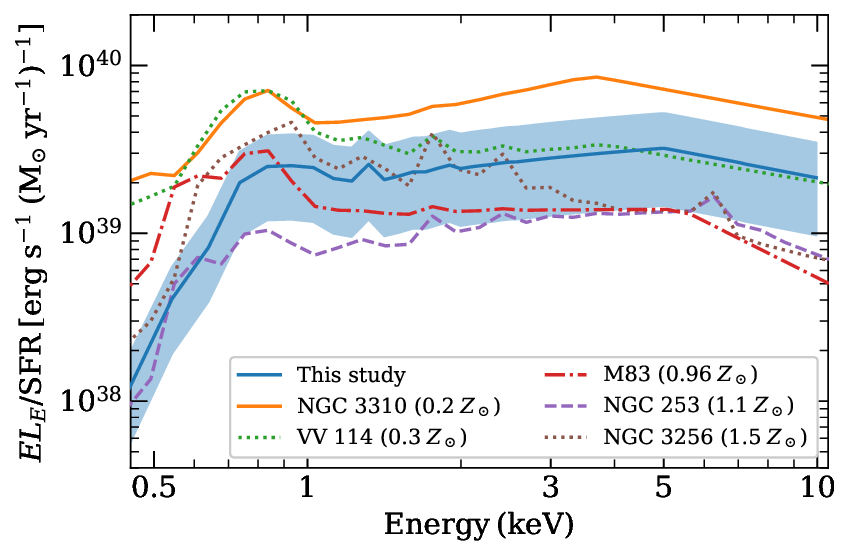}
	\caption{The best single-fit spectral shape found for the semi-independent model, normalized to SFR. The line shows the mean while the shaded region shows the best 68\% of fits. This is compared to the SEDs from NGC 3310, VV 114, M83, NGC 253, and NGC 3256, ordered in ascending metallicity.}
	\label{fig:spectra}
\end{figure}

As demonstrated in the previous section, our modeling procedure provides acceptable fits for both our fully independent and semi-independent models. Both models result in similar declines in $\gamma$ (i.e., $L_{\rm X}/M_\star$) with increasing stellar age (upper-left panel of Figure~\ref{fig:LxM}), and our fully-independent model prefers a scenario in which the \xray\ SED shape (probed by $\omega$) transitions from containing significant contributions from both hot gas and XRBs for stellar ages $<$10~Myr to an SED dominated by XRBs alone for $>$10~Myr populations (see lower-left and right panels of Figure~\ref{fig:LxM}). Our choice to define $\gamma$ in the rest-frame 2\==10~keV bandpass ensures that the resulting evolution with stellar age probes XRB evolution, since the hot gas contribution at 2\==10~keV is minimal (see, e.g., Figure~\ref{fig:omega_variance}). As such, our constraints on $\gamma$ as a function of age can be compared directly with other studies and XRB population synthesis models. \\ \indent
In the upper-left panel of Figure~\ref{fig:LxM}, we overlay the empirical results from \citet{lehmer_17}, who analyzed XRB \xray\ luminosity functions (XLFs) within subgalactic regions of M51 and correlated the observed variations of the XRB XLF with local SFHs to obtain a model for the stellar-age evolution of the XRB XLF. Due to small\==number statistics, a stellar-age-parameterized XRB XLF model was fit to the M51 data and integrated over various age bins to yield the data shown in Figure~\ref{fig:LxM}. Given this explicit parameterization, the stellar age dependent trajectory follows a continuous function and is unable to produce bin-to-bin fluctuations like those found in our results. Nonetheless, comparison of our constraints with those from M51 show a similar level of decline of $\gamma$ from 10~Myr to 10~Gyr. However, across this stellar age range, our larger number of stellar-age bins, and the independent nature of our models, result in a more detailed view of the evolution of $\gamma$ as a function of age. Furthermore, for the youngest age bin at 0\==10~Myr, we find a value of $\gamma$ that is $\sim$1 order of magnitude larger than that from \citet{lehmer_17}. This could potentially be explained by the weak constraints on this age bin in M51 and the explicit parameterization used in their XLF model. It could also be explained by comparing the metallicities of the youngest populations of stars in each sample. The median metallicity of our sample is $Z/Z_\odot \approx 0.57$, which is significantly lower than that of M51 ($Z/Z_\odot = 1.5$\==$2.5$; e.g., \citealp{Moustakas_2010}). Recent studies have established that $L_{\rm X}$(HMXB)/SFR declines with increasing metallicity \citep[see, e.g.,][]{Basu_Zych_2013b,basu_16,brorby_14,brorby_16,douna,fornasini_19,fornasini_20,kovlakas_20,lehmer_21}, and this may play some role here. \\ \indent
To provide a more physical interpretation of our results, we also displayed in the upper-left panel of Figure~\ref{fig:LxM} the XRB population synthesis model from \citet{fragos_13a}, appropriate for a solar-metallicity population. The \citet{fragos_13a} model follows a similar decline of $\gamma$ versus age as our data. Of particular interest is the abrupt order-of-magnitude decline in $\gamma$ above $\approx$10~Myr, seen in both the \citet{fragos_13a} model and our results. This decline, which is also observed in the theoretical models of \citet{linden_10}, is expected to be due to exhausting of the most luminous ULX population, typically Roche-lobe overflow HMXBs from low-metallicity stars. These binaries go through a stable common envelope phase with the compact object and an unevolved donor star, peaking in luminosity between 6 and 13 Myr, rapidly decaying after 20 Myr. We note, however, that more recent population synthesis models (e.g. \citet{wiktorowicz_17,wiktorowicz_19,Fragos_2015} that include prescriptions for beaming and a larger diversity of ULX accretor types (e.g., neutron stars), predict that ULX activity in star-forming galaxies begins just before $\approx$10~Myr, peaks between 10\==100~Myr, and strongly declines only after $\approx$100~Myr. However, it is possible that these results come from discrete time-binning effects, which also affects the timescales we investigate in this study, motivating the use of continuous models in further investigations (see Section \ref{sub:parameterized} for an example). \\ \indent
While observational constraints on the detailed formation histories associated with ULXs on timescales of $\sim$10~Myr are difficult to extract empirically, some empirical constraints on the evolution of bright binaries exist for the Small Magellanic Cloud, M31, and M33 where detailed SFHs can be extracted and correlated with X-ray point-source populations \citep[see, e.g.,][]{garofali_18,antoniou_19,lazzarini_21}. In these studies, bright \xray\ binaries are observed with an efficiency (i.e., per stellar mass) that appears to be high and approximately flat over 5\==40~Myr, before declining at ages $\simgt$50~Myr. However, the regions studied in these galaxies do not contain ULXs, which typically dominate the \xray\ power output of galaxies with SFRs $\simgt$1~$M_\odot$~yr$^{-1}$ \citep[e.g.,][]{lehmer_19,lehmer_21,kovlakas_20}, and are expected to peak at younger ages than lower-luminosity HMXBs \citep[e.g.,][]{linden_10}. We believe that given the SFR of our sample we are able to capture a more complete sampling of the HMXB XLF, which includes ULXs that would be expected to be present in higher SFR populations. This is not to say that our sample is dominated by ULXs, but that they are more represented in our sample than in low-SFR studies, such as the Magellanic Cloud, M31 and M33 studies discussed above. \\ \indent
Moving to older populations, we find an abrupt decline in $\gamma$ between the 0.1\==0.3~Gyr and 0.3\==1~Gyr age bins. The latter age range corresponds to main-sequence lifetimes of 2.5\==4~$M_\odot$ stars, an intermediate mass range for XRB donor stars that represents a transition between HMXBs and LMXBs. Following this decline, $\gamma$ appears to flatten to a value near $\log L_{\rm X}/M_\star \approx 10^{29}$~ergs~s$^{-1}$~$M_\odot^{-1}$ for populations with ages of 1\==10~Gyr. This behavior differs from that observed in the \citet{fragos_13a} population synthesis models, which predict a rapid decline in $\gamma$ between 1\==13~Gyr. The most notable discrepancies occur for the 0.3\==3~Gyr population, in which the \citet{fragos_13a} model is elevated compared to our constraints by nearly an order of magnitude. However, these intermediate mass stars are expected to go through a short-lived high-accretion state before becoming more traditional LMXBs (after going through enough mass-loss). This stage of binary evolution is difficult to capture and may cause synthesis models to inadequately describe this age range. Going forward, these results from 0\==3~Gyr can be used to better constrain binary evolution models. While some attempts have been made to isolate empirically the age-dependence of the $L_{\rm X}$(LMXB)/$M_\star$ relation within galaxies \citep[see, e.g.,][]{lehmer_07,lehmer_14,lehmer_20,kim_10,zhang}, these efforts have yielded inconclusive results and have focused primarily on the age range covered by our single age bin at 3\==10~Gyr. The M51 study by \citet{lehmer_17} provides large uncertainties on the constraints in this age range, but is consistent with our findings. \\ \indent
Regarding evolution of the \xray\ spectral shape with age (see lower left of Figure \ref{fig:LxM}), in Section~\ref{sub:fit} we showed that the 14-parameter fully independent model provides a marginal improvement to the fit to the data compared to the 8-parameter semi-independent model (based on $P_{\rm null}$ comparisons). This result, combined with prior knowledge of \xray\ SEDs of local galaxies, suggests that some age-dependency of the spectral shape is likely. In fact, the constraint on $\omega$ versus age for the fully independent model shows a strong preference for a hardening of the \xray\ SED with increasing age (see right side of Figure \ref{fig:LxM}), consistent with a rapid decline in the hot gas contribution for stellar ages above 10~Myr. Since our sample is composed of actively star-forming galaxies, it is expected that the hot gas emission component would be associated with the young population, and not the older population, which is relevant for massive elliptical galaxies \citep[e.g.,][]{boroson_2011,kim_2013b,goulding_2016,forbes_2017}. Hot gas emission in local star-forming galaxies has been observed to scale with SFR \citep[see, e.g.,][]{grimes_2005,owen_2009,mineo_12b} and has its strongest association with the most massive, and therefore youngest, stars \citep[e.g.,][]{strickland_2002,kuntz_2010,kavanagh_2020}, and our results are consistent with these observations. \\ \indent
In Figure \ref{fig:spectra}, we show our best-fit SFR-normalized spectrum of $\omega=0.78$ (from the semi-independent model) as a solid blue curve and shaded region (representing the full range of the top 68\% of fits, in terms of $C$). For comparison purposes, we show the spectral fits from other relatively nearby star-forming galaxies: VV 114 (\citealp{garofali_20}), NGC 253 (\citealp{Wik_2014}), M83 (\citealp{Yukita_2016}), NGC 3256, and NGC 3310 (\citealp{lehmer_15}). These galaxies have excellent spectral constraints across 0.3\==30~keV from \chandra\ or \xmm\ combined with \nustar\ data, allowing for decomposition of hot gas and XRB contribution across that bandpass. Additionally, they were selected to demonstrate the range of spectral shapes observed for a variety of metallicities: 0.3, 1.1, 0.96, 1.5, and 0.2~$Z/Z_\odot$, respectively. The single best-fit spectral shape that we find is largely consistent with that of NGC 3256 and VV 114, although the general shape is similar to other SEDs as well. Both NGC 3256 and VV 114 are starburst galaxies, so the similarity in SEDs reinforces the observed elevated SFR of our sample, noted earlier in Section \ref{sub:phys_deriv}.\\ \indent
For comparison with more commonly used scaling relations we calculated values of the equivalent relations implied by our model (listed in Table \ref{table:scaling_relations} alongside 2\==10 keV scaling relations). For the HMXB scaling with SFR \citep[e.g.,][]{ranalli_03,mineo_12a,lehmer_19} derived from the three youngest $\gamma$ values and HMXB component spectra in our model, we multiplied by $10^8$ yr in order to convert from $M_\star$ to SFR as the scaling factor, and further scaled using the XRB component of the spectra (shown in the top of Figure \ref{fig:omega_variance}) from 0.5$-$8 keV ($L^{\rm HMXB}_{0.5-8 \; \rm{keV}}/L^{\rm total}_{2-10 \; \rm{keV}}$). We calculate $\log L_{0.5-8 \; \rm keV}$[HMXB]/SFR = $39.68^{+0.36}_{-0.66}$, which falls right between the modeled values of $39.80$ and $39.64$ from \citet{lehmer_21}, which correspond to metallicities of 0.51 and 0.81 $Z/Z_\odot$, matching with the median metallicity of our sample ($0.57 \; Z/Z_\odot$). This also agrees with the full observed sample value of $39.71^{+0.14}_{-0.09}$ from \citet{lehmer_19}. \\ \indent
For the hot gas emission of our spectra (shown in the top of Figure \ref{fig:omega_variance}) we followed a similar process as for the HMXBs, as the youngest age range is where hot gas is expected to be found. The final numbers were scaled to the hot gas 0.5$-$2 keV component instead of the previous values for HMXBs. We determined how the hot gas \xray\ emission scales with SFR. We calculate a value $L^{\rm gas}_{0.5-2 \; \rm keV}/\rm{SFR} = (9.15^{+11.77}_{-7.13}) \times 10^{38}$~erg~s$^{-1}$~($M_\odot$~yr$^{-1}$)$^{-1}$, which is consistent with the value of $7.75 \pm 0.3 \times 10^{38}$~erg~s$^{-1}$~($M_\odot$~yr$^{-1}$)$^{-1}$ reported in \citet{mineo_12b} (after correcting for different IMF assumptions). This value was already noted to have significant spread (a factor of $\sim5$) through multiple studies, so there might be underlying factors which need to be accounted. \\ \indent
% for Owen & Warwick they find a range of 5.39e38 to 2e39 after converting from 0.3-1 keV (their soft band) to 0.5-2 keV using our own gas component spectra
%
For LMXBs, we calculated $\log (L_{0.5-8 \; \rm{keV}}$[LMXB]/$M_\star) = 29.13^{+0.43}_{-0.32}$ using the oldest two age-bins and scaling to the 0.5$-$8 keV XRB spectral component. Because this scaling relation remains proportional to $M_\star$ the multiplier of $10^8$ yr is not used. By using the oldest two age-bins we limit LMXBs to be older than 1~Gyr, which might not be properly accounting for more intermediate mass stars reaching Roche lobe overflow earlier than 1~Gyr (2-5 $M_\odot$), so this value can be considered a lower-limit. This is consistent with values reported in previous studies of elliptical galaxies, including $\sim$28.98 from \citet{zhang} and $28.86^{+0.07}_{-0.08}$ from \citet{lehmer_20}. \\ \indent
\begin{table}
\centering
 \begin{tabular}{ l r }
 \hline
 \hline
 \multicolumn{2}{c}{Scaling Relations} \\
 \hline
 \hline
 \rule{0pt}{1ex} log($L^{\rm gas}_{0.5-2 \; \rm keV}/\rm{SFR} \; [\rm{erg \; s^{-1} \; (M_\odot \; yr^{-1})^{-1}}]$) & $38.96^{+0.36}_{-0.66}$ \\
 \rule{0pt}{4ex} log($L^{\rm LMXB}_{0.5-8 \; \rm keV}/\rm{M_\star} \; [\rm{erg \; s^{-1} \; M^{-1}_\odot}]$) & $29.13^{+0.43}_{-0.32}$ \\
 \rule{0pt}{4ex}  log($L^{\rm LMXB}_{2-10 \; \rm keV}/\rm{M_\star} \; [\rm{erg \; s^{-1} \; M^{-1}_\odot}]$) & $29.03^{+0.43}_{-0.32}$ \\
 \rule{0pt}{4ex} log($L^{\rm HMXB}_{0.5-8 \; \rm keV}/\rm{SFR} \; [\rm{erg \; s^{-1} \; (M_\odot \; yr^{-1})^{-1}}]$) & $39.68^{+0.36}_{-0.66}$ \\
 \rule{0pt}{4ex} log($L^{\rm HMXB}_{2-10 \; \rm keV}/\rm{SFR} \; [\rm{erg \; s^{-1} \; (M_\odot \; yr^{-1})^{-1}}]$) & $39.57^{+0.36}_{-0.66}$ \\
 \hline
 \hline
\end{tabular}
\caption{Scaling relations, derived from the fully independent model in order to fully account for changes in the spectral shape, for the \xray\ luminosity of hot gas and HMXBs scaled to SFR, and LMXBs scaled to $\rm{M}_\star$. The hot gas and HMXB scaling relations are based on the youngest three age-bins (less than 100 Myr) while the LMXB scaling relation is calculated using the oldest two age bins (greater than 1 Gyr).}
\label{table:scaling_relations}
\end{table}
\subsection{Parameterized fits} \label{sub:parameterized}
\begin{figure}
	\centering
	\includegraphics[width=8.75cm]{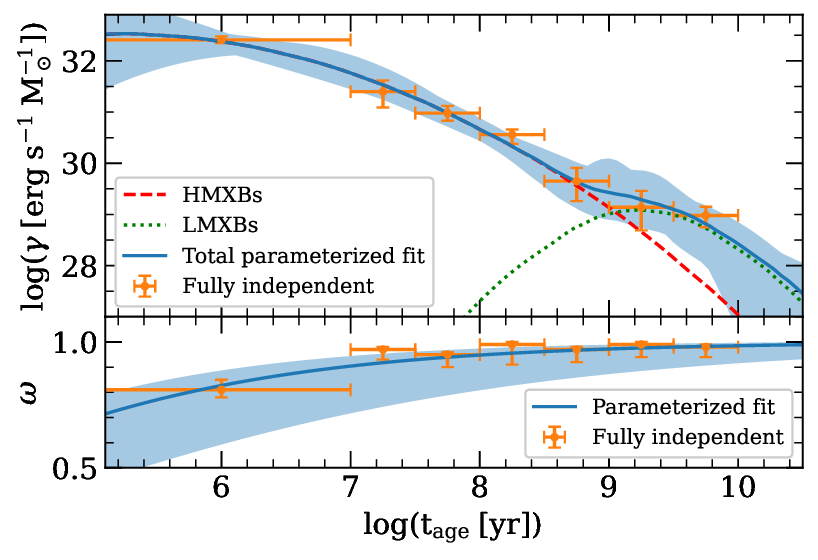}
	\caption{The parameterized fits for $\gamma$ and $\omega$ presented alongside the results from the fully-independent model (orange points). \textit{Top}: The parameterized formula for $\gamma$ can be decomposed into the contributions from HMXBs and LMXBs (shown as red dashed and green dotted lines respectively). The constraints become very weak above 9.75 Gyr because that is the assumed age of the oldest age-bin in this study. The shaded region shows the uncertainty as derived from the top 68\% of the fully independent model results ("top" is defined by having the lowest $C$ value). \textit{Bottom}: The parameterized fit of $\omega$ is shown with a similar uncertainty region reflecting the top 68\% of results. }
	\label{fig:para_fits}
\end{figure}

An important step in the methodology of this study was assuming that the SFR will be constant across each of the seven age-bins, with immediate changes occurring between any two age-bins. This allows for a simple model to be created (seven distinct values for $\gamma$ and $\omega$), but obviously it is not necessarily physically motivated. In order to get an idea for what a continuous model would produce we created a new model that parameterizes the values of $\gamma$ and $\omega$ as continuous functions of time, and used it to fit the results from our fully independent model. \\ \indent
To describe the overall process of XRB populations aging we created a parameterized model of $\gamma$ with two distinct components: one for HMXBs and one for LMXBs. This does leave out hot gas emission, but, as our evolution of $\omega$ indicates, the hot gas is assumed to be part of the HMXB curve. To prevent these curves from becoming non-physical (i.e. taking on values that are too large) at certain ages we used quadratic curves with the restriction of having negative curvature (represented by $a_0$ and $a_1$). \\ \indent
\begin{equation}
\gamma (\rm{t_{age}}) = \gamma_{\rm HMXB} (\rm{t_{age}}) + \gamma_{\rm LMXB} (\rm{t_{age}})
\end{equation}
\begin{equation}
\gamma (\rm{t_{age}}) = \mathit{a_0(\tau-b_0)^2+c_0 + a_1 (\tau-b_1)^2+c_1}
\end{equation}
In these equations $\tau$ is simply used as an abbreviation for $\rm{log(t_{age})}$. This resulted in a single curve that can easily be decomposed into two components. One additional restriction was placed on the parameterization: because we do not have age data exceeding 9.75 Gyr (the assumed age of the oldest age bin) we adopted a prior in which we limited the value of $b_1$ (the age where LMXB emission would peak) to be less than 9.75 Gyr. \\ \indent
We also created a two-parameter fit for $\omega$. Because $\omega$ is bound between 0 and 1, and should be 0 at extremely low ages (before XRBs activate) and 1 at extremely high ages (when only XRBs remain) we chose to use a Gompertz function that is bound between 0 at negative infinity and 1 at positive infinity. The parameters used are $b_\omega$ and $c_\omega$, which define the "midpoint" of the curve and the rate of increase at the midpoint. The parameterization thus follows the explicit form:
\begin{equation}
\omega (\rm{t_{age}}) = \rm{exp}(-\mathit{e^{b_\omega-c_\omega \tau}}).
\end{equation}
In total this results in a parameterized model with two separate equations ($\gamma$ and $\omega$) each in terms of age, with a total of 8 parameters. The results for this parameterization are shown in Figure \ref{fig:para_fits} and listed in Table \ref{table:parameterized}. The fits visually appear to fit the data well, but some of the parameters (particularly those for the LMXB component of the $\gamma$ fit) seem to have wide ranges of uncertainty. This reflects a high degree of degeneracy and covariances in the model, which is most likely caused by the fact that the LMXB component is primarily constrained by two data points and our adopted prior on $b_1$. A similar degree of covariance can be seen in the HMXB component, where the emission can peak anywhere below 3 Myr (shown as $b_0$), coming from the lack of any age constraints below 10 Myr. Additionally, this peak-age is impacted by the emission of hot gas, as it is expected to play a significant role in the younger stellar populations. These parameterized results are weakly constrained given our current data set, but can be utilized in future studies to examine how these scaling relations vary with other physical parameters. For example, metallicity has been identified as an important factor influencing HMXB formation, so the parameters for the HMXB model could be modified to include a metallicity dependence.
\begin{table}
\centering
 \begin{tabular}{ l r }
 \hline
 \hline
 \multicolumn{2}{c}{Parameterized Model Results} \\
 \hline
 \hline
 $a_0 \; [\rm{erg \; s^{-1} \; M_\odot^{-1} \; log(yr)^{-1}}]$ & $-0.24^{+0.19}_{-0.35}$ \\
 $b_0 \; [\rm{log(yr)^{-1}}]$ & $5.23^{+1.10}_{-5.23}$ \\
 $c_0 \; [\rm{erg \; s^{-1} \; M_\odot^{-1}}]$ & $32.54^{+2.61}_{-0.40}$ \\
 $a_1 \; [\rm{erg \; s^{-1} \; M_\odot^{-1} \; log(yr)^{-1}}]$ & $-1.21^{+0.57}_{-4.59}$ \\
 $b_1 \; [\rm{log(yr)^{-1}}]$ & $9.32^{+0.17}_{-0.32}$ \\
 $c_1 \; [\rm{erg \; s^{-1} \; M_\odot^{-1}}]$ & $29.09^{+0.91}_{-1.08}$ \\
 \hline
 $b_\omega $ & $2.15^{+0.04}_{-0.08}$ \\
 $c_\omega \; [\rm{log(yr)^{-1}}]$ & $0.63^{+0.06}_{-0.18}$ \\
 \hline
 \hline
\end{tabular}
\caption{The values for the parameters used to fit the full independent model results (see Section \ref{sub:parameterized} for the equations used). Several of the values are consistent with limits set on the parameters, and show high degrees of covariance, especially the LMXB component of the $\gamma$ model ($a_1, b_1, c_1$).}
\label{table:parameterized}
\end{table}
\subsection{Additional physical dependencies} \label{sub:subs}
\begin{figure}
	\centering
	\includegraphics[width=8.75cm]{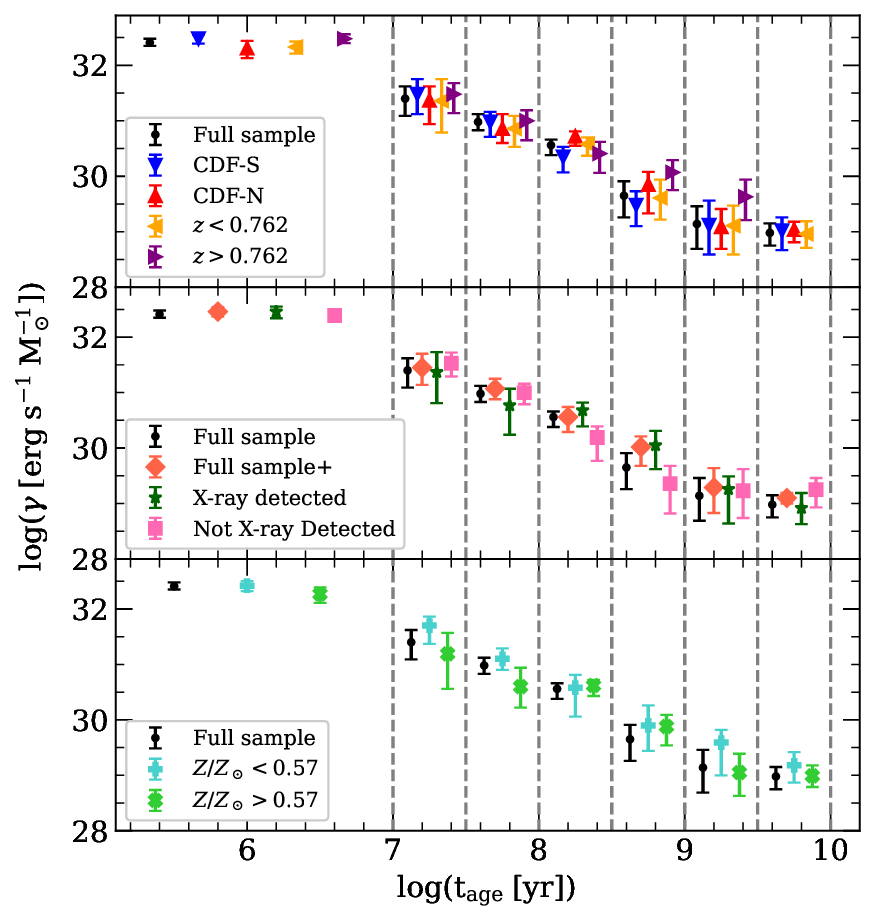}
	\caption{The evolution of $\gamma_i$ for four different splits of the sample, resulting in eight different subsamples. The seven age bins are separated by vertical dashed lines to show the subsamples in a more readable manner. \textit{Top}: Moving from left to right within each age-bin the subsample splits are as follows: the complete sample (black points); CDF-S / CDF-N (blue downwards triangles/ red upwards triangles); and low redshift / high redshift (orange leftwards triangles/ purple rightwards triangles). \textit{Middle}: Moving from left to right within each age-bin the subsample splits are as follows: the full sample (black points); the full sample plus 46 galaxies that were originally classified as normal galaxies, but not included in our full sample due to an overlap in properties with known AGN (orange diamonds); the 47 \xray\ detected galaxies (dark green stars); the 294 non-\xray\ detected galaxies (pink squares). \textit{Bottom}: Moving from left to right within each age-bin the subsample splits are as follows: the complete sample (black points); low metallicity (light blue $+$ symbols); high metallicity (light green $\times$ symbols).}
	\label{fig:subsamples}
\end{figure}

In order to probe whether additional unmodeled dependencies may be relevant, we tested how the results of the fully independent model would vary if the sample were split into various subsamples. One of the original motivations for creating a physically-dependent model is that it can be properly extrapolated out to high redshifts and still return accurate values. We tested splitting the sample up in four different ways: by field (CDF-S vs CDF-N), by redshift, by \xray\ detection, and by metallicity. \\ \indent
The results from the fully independent model for each of the eight subsamples are shown in Figure \ref{fig:subsamples}, along with an extended sample discussed later in this section. The presentation is changed compared to Figure \ref{fig:LxM} in order to have the points be more readable. Each of the seven age-bins are now separated by vertical gray dashed lines, with data points within each age-bin (subsamples and the complete sample). Each of the data points within an age-bin reflect the value for the entirety of that age-bin, however the horizontal error bars have been excluded and the points have been horizontally offset from each other for ease of viewing. \\ \indent
To quantify these subsample comparisons, we define a parameter $\epsilon$ that measures the average ratio of $\gamma$ between two subsamples ($A$ and $B$) across all seven age-bins (or six age-bins in the case of the redshift split since the highest age-bin is unavailable for high redshifts): \\ \indent
\begin{equation}
\epsilon = \frac{1}{N} \sum_{i=1}^{N} \frac{\gamma_{i,A}}{\gamma_{i,B}}
\end{equation}
The value of $\epsilon$ and its marginalized probability distribution are calculated by propagating the MC results described in Section~\ref{sub:fit}. Below, we quote median and 16--84\% confidence ranges for $\epsilon$.

\begin{figure}
	\centering
	\includegraphics[width=8.75cm]{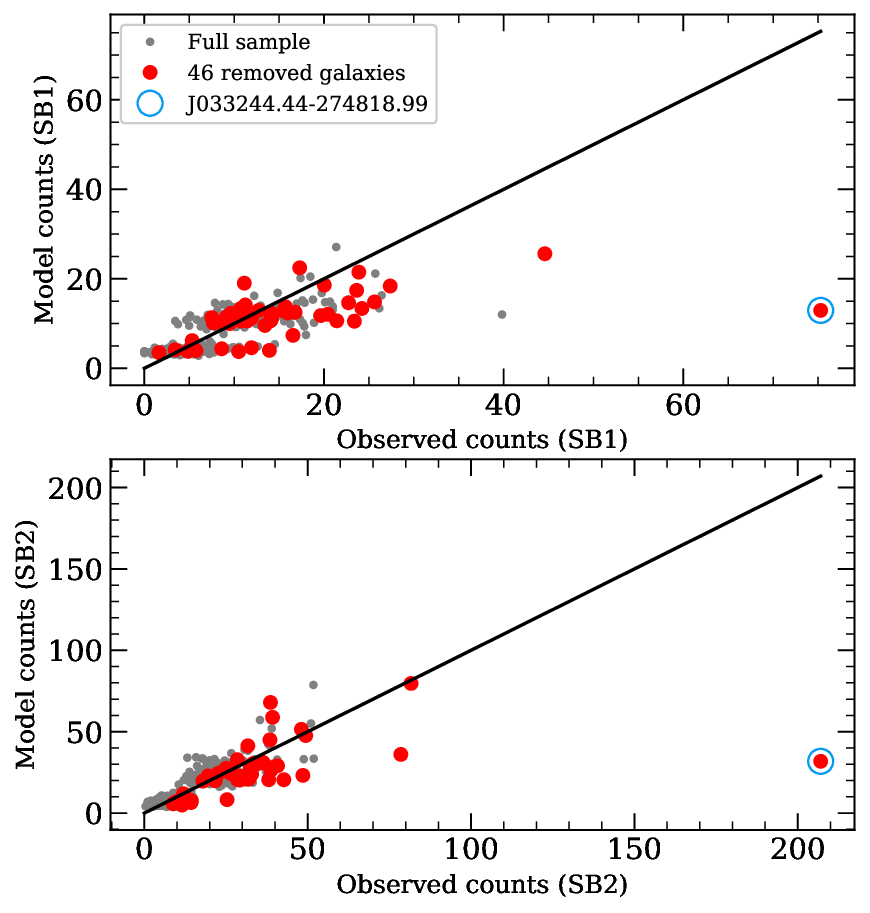}
	\caption{A comparison of model counts versus observed counts in each the SB1 (\textit{upper}) and SB2 (\textit{lower}). The 344 galaxies of the full sample are shown as small gray points with the 46 removed galaxies (for possibly being undeteced AGN, see Section \ref{sec:sample_select}) are shown are red points. A one-to-one black line is shown to display where points would lie when the model and observed counts agree. One significant outlier source (J033244.44-274818.99) is highlighted with a light blue circle.}
	\label{fig:removed_counts}
\end{figure}

In the top portion of Figure \ref{fig:subsamples}, we present results for four of the subsamples that are expected to be consistent with those obtained from the full sample of 344 galaxies. The left-most point within each bin shows the full sample as a black point. The next two points show the results for the CDF-S (blue downwards triangles) and CDF-N (red upwards triangles), which remain entirely consistent with each other across all age-bins, and have their $C_{\rm min}$ values listed in Table \ref{table:parameters}. For this subsample split (measuring $\epsilon$ as the ratio of CDF-S to CDF-N), we obtain $\epsilon = 0.91^{+0.45}_{-0.30}$, showing that there is no significant difference in results between fields. \\ \indent
The next two points in the top panel of Figure \ref{fig:subsamples} (orange leftwards triangles and purple rightwards triangles) show the redshift split: low redshift galaxies first and high redshift galaxies following them. We divided the sample in half by redshift at $z=0.762$ (each subsample containing 172 galaxies). These are also consistent with each other across all age ranges, indicating that this model does not have any detectable redshift dependency. For this subsample split (measuring the ratio of the low-$z$ subsample to the high-$z$ subsample), we calculate $\epsilon = 0.64^{+0.39}_{-0.24}$, a value that is somewhat lower than unity, but still consistent with no significant difference between the two subsamples. A value of $\epsilon < 1$, could potentially be expected due to a small difference between the metallicity of the two redshift-split subsamples, as we expect metallicity to be lower at higher redshifts. Indeed, we estimate sample median values of $Z=0.52 Z_\odot$ for the high-$z$ subsample and $Z=0.60 Z_\odot$ for the low-$z$ subsample. \\ \indent
The middle panel of Figure \ref{fig:subsamples} displays three different subsamples: the first (orange diamonds) is an extended sample (full sample+), which includes the full sample plus the additional 46 normal-galaxy-classified sources that were conservatively removed during the sample selection process as described in Section~\ref{sec:sample_select}; the second subsample consists of only the 47 \xray\ detected galaxies in our sample (dark green stars, see Section \ref{sub:xray} for further details); and the third subsample contains the 294 non-\xray\ detected galaxies in our sample (pink squares). \\ \indent
Overall the "full sample+" of 390 galaxies produces a similar result to the standard full sample in terms $\gamma$ evolution over stellar age. The measured $\epsilon$ value for the ratio of full sample+ to the full sample is $1.33^{+0.58}_{-0.41}$ which is again consistent with unity. To see exactly how the model fits these sources we have plotted the model counts versus the observed counts in Figure \ref{fig:removed_counts}. The 46 removed sources are shown as the red points, which do appear to be broadly underestimated by the model (demonstrated by the side of the one-to-one line that they lie along). One such source in particular (J033244.44-274818.99) is found to have significantly higher observed counts than the model predicts in both SB1 and SB2 (highlighted with a blue circle around it). This source is not classified as an AGN in any other catalog, but given its significant outlier status with respect to our normal-galaxy model, we consider this source to be an AGN candidate. As we predicted earlier, a majority of these removed galaxies are consistent with their original normal galaxy classification. Our results show that our model may be useful in future studies to identify potential AGN candidates. \\ \indent
The \xray\ detected subsample does appear to have insignificantly elevated $\gamma$ values compared to the non-\xray\ detected subsample, however this is expected, since these galaxies are specifically selected for their \xray\ brightness. Additionally, this subset has a significantly smaller sample size relative to the full sample (47 galaxies compared to the full 344 galaxy sample) and is subject to potential stochastic effects. The overall evolution of $\gamma$ follows similar trends for each subsample and the overall sample. For this subsample split (measuring $\epsilon$ as the ratio of \xray\ detected galaxies to undetected galaxies), we calculate that $\epsilon = 1.15^{+0.61}_{-0.40}$, which is consistent with no significant difference. \\ \indent
The final split displayed in the lower portion of Figure \ref{fig:subsamples} is between the low and high metallicity galaxies (light blue $+$ and light green $\times$ symbol, respectively; 172 galaxies in each subsample). Again, the full sample is shown as black points for comparison. The metallicity for each galaxy was calculated following the same method outlined in Section \ref{sub:xray} and a value of $Z/Z_\odot=0.57$ was used to split the sample in half, with the low-metallicity subsample having a mean metallicity of $Z/Z_\odot=0.45$ and the high-metallicity subsample having a mean of $Z/Z_\odot=0.78$. This subsample split does result in significant differences in numerous age-bins, from the younger through the older stellar populations. This difference is entirely expected, as discussed in Section \ref{sec:intro}. Decreased metallicity has been shown to correlate with higher $L_{\rm X}$/SFR values, indicating a greater/more luminous presence of HMXBs. We calculate that, for a ratio of low-metallicity to high-metallicity, $\epsilon=2.63^{+1.96}_{-1.25}$. While this value does have significant uncertainty, it suggests that the low-$Z$ subsample is $\approx$1.5--4.5 times more luminous per unit stellar mass than the high-$Z$ subsample. Separate $\epsilon$ values for HMXBs and LMXBs can be estimated by selecting the four youngest age-bins for HMXBs and the three oldest age-bins for LMXBs. When calculated this way we find that $\epsilon_{\rm{HMXB}}=2.94^{+2.19}_{-1.4}$ and $\epsilon_{\rm{LMXB}}=2.23^{+1.66}_{-1.06}$, confirming that metallicity plays a significant role in the evolution of both LMXBs and HMXBs.  The elevated LMXB emission in low-metallicity galaxies is expected from in-situ star-formation activity \citep[e.g.][]{fragos_13a}, but differs from that observed in globular cluster LMXBs, which are expected to form dynamically and have been shown to have more luminous LMXB populations in high-metallicity environments (e.g. \citealp{kundu_2002,ivanova_2012,kim_2013a}).  Our results therefore provide suggestive evidence that the LMXB populations in these star-forming galaxies may be dominated by the in-situ formation pathway.
While these results will not be investigated further in this study, they serve as motivation for future studies on parameterizing metallicity as a factor of XRB evolution. We note that when comparing the $\gamma$ values for the full sample with those of the metallicity-split subsamples, it is apparent that the low-metallicity galaxies drive the full sample result at ages $<$0.1~Gyr, and the high-metallicity galaxies drive the full-sample result at older ages. Therefore, our full sample result is not expected to be appropriate for a single mean metallicity. \\ \indent
The metallicity subsample split is expected to show some difference, as tension between the two subsamples is visible in Figure \ref{fig:subsamples} and has been detected in previous studies (see Section \ref{sec:intro}). This effect is expected and has been observed in other studies before. Based on the $L_{\rm X}$-SFR-$Z$ relation from \cite{lehmer_21} we would expect to find an $\epsilon$ ratio of $2.00^{+1.31}_{-0.80}$, which is consistent with our value. While this study will not be further pursuing the effects of metallicity on XRB (particularly HMXB) luminosity, we have outlined a potential method to do so (see Section \ref{sub:parameterized}) and have found evidence to continue pursuing this in further studies. \\ \indent
\section{Summary} \label{sec:summary}
Using a sample of 344 normal, actively star-forming galaxies in the CDFs that span a redshift range of $z =$~0--3.5, we have investigated how the X-ray luminosity scaling with stellar mass ($L_{\rm X}/M_\star$) and X-ray spectral shape depend on stellar age. We have conducted SED fitting of FUV-to-FIR photometric data to constrain SFHs for each of our galaxies, and we have correlated these SFHs with constraints on the X-ray emission from the galaxies using the ultradeep \chandra\ data in the CDFs. Our analyses indicate significant changes in both $L_{\rm X}/M_\star$ and X-ray spectral shape as stellar populations age. Specific findings from our study include:
\begin{enumerate}
    \item A factor of $\sim$1000 decline in $L_{\rm X}/M_\star$ over the age range of 10 Myr--10 Gyr is found, consistent with theoretical models and studies of a small number of local galaxies. Our analysis provides unique statistical constraints based on a large sample of deep-field galaxies, quantifying the continuous transition from luminous HMXBs associated with young populations to LMXBs associated with older populations. 
    \item We find that the \xray\ SED shape becomes harder at ages $\simgt$10~Myr, which is consistent with a scenario where the SED contains significant contributions from both hot gas and XRBs associated with $\simlt$10~Myr populations and primarily XRBs at $\simgt$10~Myr. The interpretation of this result, however, requires more robust constraints on the variation of \xray\ spectral shape with age.
    \item Global scaling relations $L_{\rm X}$(gas)/SFR, $L_{\rm X}$(HMXB)/SFR, and $L_{\rm X}$(LMXB)/$M_\star$ are derived by selecting parameter fit results from subsets of age-ranges (younger populations for hot gas and HMXBs, older populations for LMXBs). The scaling relationships we derive are consistent with those found in previous studies of galaxies in the local universe.
    \item We present parameterizations of $L_{\rm X}/M_\star$ and spectral shape as a function of stellar age assuming separate contributions from HMXBs and LMXBs. Our parameterizations prefer a continuous decline of HMXB emission with increasing age, and an onset of an LMXB emission component at $\simgt$1~Gyr. This modeling characterizes well an overall observed flattening of $L_{\rm X}/M_\star$ that is observed for ages $\simgt$1~Gyr.
    \item We split the full sample of 344 galaxies into subsamples based on observational field (CDF-N vs CDF-S), redshift, X-ray brightness, and metallicity to search for further dependencies. We find no significant dependence on observational field, redshift, or \xray\ brightness. However, we find that the low-metallicity half of our galaxy sample shows enhanced levels of $L_{\rm X}/M_\star$ compared to the high-metallicity half. This metallicity dependence is predicted theoretically and has been observed in other studies. These results warrant future investigations aimed at quantifying how $L_{\rm X}/M_\star$ and SED shape depend on both age and metallicity.
\end{enumerate}
\begin{acknowledgments}

We thank the anonymous referee for their helpful comments, which have improved the quality of this paper. We gratefully acknowledge financial support from the NASA Astrophysics Data Analysis Program (ADAP) grant 80NSSC20K0444 (W.G.,B.D.L.,K.D.,R.T.E.) and \chandra\ X-ray Center (CXC) grant AR9-20008X (W.G.,B.D.L.). W.N.B. acknowledges support from the V.M. Willaman Endowment. T.F. acknowledges support from the Swiss National Science Foundation Professorship grant (project number PP00P2 176868). K.G. is supported by an appointment to the NASA Postdoctoral Program at NASA Goddard Space Flight Center, administered by Universities Space Research Association under contract with NASA. B.L. acknowledges financial support from the National Natural Science Foundation of China grant 11991053, China Manned Space Project grants NO. CMS-CSST-2021-A05 and NO. CMS-CSST-2021-A06. Y.Q.X. acknowledges support from NSFC-12025303, 11890693, 11421303, the CAS Frontier Science Key Research Program (QYZDJ-SSW-SLH006), the K.C. Wong Education Foundation, and the science research grants from the China Manned Space Project with NO. CMS-CSST-2021-A06. \\ \indent
This work is based on observations taken by the CANDELS Multi-Cycle Treasury Program with the NASA/ESA HST, which is operated by the Association of Universities for Research in Astronomy, Inc., under NASA contract NAS5-26555. This work has made use of the Rainbow Cosmological Surveys Database, which is operated by the Centro de Astrobiologia (CAB/INTA), partnered with the University of California Observatories at Santa Cruz (UCO/Lick, UCSC), and the Arkansas High Performance Computing Center, which is funded through multiple National Science Foundation grants and the Arkansas Economic Development Commission. The material is based upon work supported by NASA under award number 80GSFC21M0002.

\end{acknowledgments}

\bibliographystyle{aasjournal}
\bibliography{mybib}

\end{document}